\documentclass[fleqn,usenatbib]{mnras}

\usepackage[T1]{fontenc}

\DeclareRobustCommand{\VAN}[3]{#2}
\let\VANthebibliography\thebibliography
\def\thebibliography{\DeclareRobustCommand{\VAN}[3]{##3}\VANthebibliography}

\usepackage{graphicx}	
\usepackage{amsmath}	
\usepackage{amssymb}	
\usepackage{xcolor}
\usepackage{caption}
\usepackage{subcaption}
\usepackage{textcomp}
\usepackage{float}
\usepackage{bm}
\usepackage{enumerate}

\newcommand{\hcfn}{HC$_5$N}
\newcommand{\amm}{NH$_3$}
\newcommand{\hh}{H$_2$}
\newcommand{\nh}{$N(\mathrm{H}_2)$}
\newcommand{\nbg}{$N_{\text{bg}}$}
\newcommand{\vlsr}{$v_\mathrm{LSR}$}
\newcommand{\kms}{km\,s$^{-1}$}
\newcommand{\kmspc}{km\,s$^{-1}$~pc$^{-1}$}
\newcommand{\cc}{cm$^{-3}$}
\newcommand{\tex}{$T_{\text{ex}}$}
\newcommand{\Gi}{$G_1$}
\newcommand{\Gii}{$G_2$}
\newcommand{\Giii}{$G_3$}

\newcommand\mub{\bm{\mu}}
\newcommand\ab{\bm{a}}
\newcommand\sigmab{\bm{\sigma}}

\title[Inflow and Fragmentation in TMC-1]{Velocity-Coherent Substructure in TMC-1: \\ Inflow and Fragmentation}

\author[Smith et al.]{Simon E.T. Smith$^{1,2}$,\thanks{E-mail: simonsmith@uvic.ca} Rachel Friesen$^{1}$, Antoine Marchal$^{3, 4}$, Jaime E. Pineda$^{5}$, Paola Caselli$^{5}$, \newauthor Michael Chun-Yuan Chen$^{6}$, Spandan Choudhury$^{5}$, James Di Francesco$^{2, 7}$, Adam Ginsburg$^{8}$, \newauthor Helen Kirk$^{7}$, Chris Matzner$^{1}$, Anna Punanova$^{9}$, Samantha Scibelli$^{10}$, Yancy Shirley$^{10}$
\\
$^{1}$David A. Dunlap Department of Astrophysics and Astronomy, 50 St George St, Toronto, ON, M5S 3H4, Canada \\
$^{2}$Department of Physics and Astronomy, University of Victoria, Victoria, BC, V8P 1A1, Canada \\
$^{3}$Canadian Institute for Theoretical Astrophysics, University of Toronto, 60 St. George Street, Toronto, ON M5S 3H8, Canada \\
$^{4}$Research School of Astronomy \& Astrophysics, Australian National University, Canberra ACT 2610 Australia \\
$^{5}$Max-Planck-Institut f\"ur extraterrestrische Physik, Giessenbachstrasse 1, 85748 Garching, Germany \\
$^{6}$Queen’s University, 99 University Ave, Kingston, ON, K7L 3N6, Canada \\
$^{7}$Herzberg Astronomy and Astrophysics, National Research Council of Canada, 5071 West Saanich Road, Victoria, BC, V9E 2E7, Canada \\
$^{8}$Department of Astronomy, University of Florida, PO Box 112055, USA \\
$^{9}$Ural Federal University, 620002, Mira st. 19, Yekaterinburg, Russia \\
$^{10}$Steward Observatory, 933 North Cherry Avenue, Tucson, AZ 85721, USA
}

\date{Accepted XXX. Received YYY; in original form ZZZ}

\pubyear{2022}

\begin{document}
\label{firstpage}
\pagerange{\pageref{firstpage}--\pageref{lastpage}}
\maketitle

\begin{abstract}

Filamentary structures have been found nearly ubiquitously in molecular clouds and yet their formation and evolution is still poorly understood.
We examine a segment of Taurus Molecular Cloud 1 (TMC-1) that appears as a single, narrow filament in continuum emission from dust.
We use the Regularized Optimization for Hyper-Spectral Analysis ({\tt ROHSA}), a Gaussian decomposition algorithm which enforces spatial coherence when fitting multiple velocity components simultaneously over a data cube.
We analyze HC$_5$N (9-8) line emission as part of the Green Bank Ammonia Survey (GAS) and identify three velocity-coherent components with {\tt ROHSA}.
The two brightest components extend the length of the filament, while the third component is fainter and clumpier. 
The brightest component has a prominent transverse velocity gradient of $2.7 \pm 0.1$\,\kmspc\ that we show to be indicative of gravitationally induced inflow.
In the second component, we identify regularly spaced emission peaks along its length. We show that the local minima between pairs of adjacent HC$_5$N peaks line up closely with submillimetre continuum emission peaks, which we argue is evidence for fragmentation along the spine of TMC-1. 
While coherent velocity components have been described as separate physical structures in other star-forming filaments, we argue that the two bright components identified in HC$_5$N emission in TMC-1 are tracing two layers in one filament: a lower density outer layer whose material is flowing under gravity towards the higher density inner layer of the filament.

\end{abstract}




\section{Introduction}
\label{sec:intro}


Star formation appears to occur predominantly within molecular cloud filaments \citep[see review by][ and references therein]{andre_2014_PPVI}.
Some studies have suggested that these dense filaments have typical widths of 0.1 pc \citep{Arouz_2011}, though more recent studies have called this `universal width' into question \citep{panopoulou_2017, panopoulou_2022, ossenkopf_2019}. Star forming filaments often have column densities of N$_{\text{H}_2}^{\text{fil}}$ $\gtrsim$ 7\,$\times$ 10$^{21}$\,cm$^{-2}$ \citep{K_nyves_2015} and dust temperatures $T_\text{d} \simeq 10$\,K.
Kinematic studies using dense gas tracing molecules have been integral in probing the internal structure and further fragmentation of these
formations \citep{Pineda2022}.

Velocity gradients have been observed across filaments \citep{fernandez_2014a, fernandez_2014b, dhabal_2018, Shimajiri_2019} and are often interpreted as accretion of fresh material. Similarly, velocity gradients have been identified along filaments \citep{friesen_2013, kirk_2013, peretto_2014} and are often interpreted as the flow of material along the structure.
Simulations show that supersonic turbulent flows \citep{vazquez_2007, hgong_2011, mgong_2015} and magnetic fields \citep{ostriker_1999, ciolek_2006} can both create  sheet-like structures of gas. It has been demonstrated that the subsequent collapse of these sheets can produce observable transverse velocity gradients \citep{CO_2014, CO_2015}. 
With high spatial and spectral resolution, some studies have identified velocity-coherent substructures that match well to overdense regions in dust continuum emission maps \citep[e.g. ][]{Hacar_2013, Hacar_2017}. Although they often overlap in projection on the sky, these features, sometimes dubbed `fibres' \citep[][]{andre_2014_PPVI}, have been described as physically distinct components of gas, intertwined along a filament, identifiable through coherent gas motions.
Conversely, it has been argued that these features do not necessarily map to physically distinct structures in position-position-position (PPP) space \citep[][]{ZamoraAviles_2017, Clarke_2018}.

The diverse kinematics of filaments have primarily been investigated using chemical tracers to reveal the movement of gas in different physical environments.
Ammonia (\amm) is formed and excited at densities $n \simeq 10^4$\,cm$^{-3}$ \citep{ho_townes_1983,densecores_book} and therefore has been used frequently to trace the densest parts of molecular clouds, including filaments. At such high densities and low temperatures, many other molecular species freeze out onto dust grains or are depleted via chemical reactions \citep{bergin_1997, walmsley_2004, Walsh_2009}. \amm\ has proven to be resilient to such depletion processes \citep{aikawa_2001, densecores_book} up until $n \simeq 10^6$\,cm$^{-3}$ \citep{walmsley_2004}, making it an ideal tracer for dense structures in filaments.

Carbon-chain species, such as CCS or HC$_n$N ($n=3, 5, 7, ...$) are also found in star-forming regions, 
both towards cold, quiescent cores \citep{snell_1981, hirota_2009, Taniguchi_2017}, as well as in warm environments around young protostars \citep{sakai_2008, sakai_2009, sakai_2013, law_2018}, and the recently identified streamers \citep[][]{Pineda2020}. In particular, the \hcfn\ $J = 9-8$ transition traces similar gas densities as the \amm\ (1,1) and (2,2) inversion transitions.
Toward cold cores, the presence of carbon-chain species has been used to argue that the gas is chemically and dynamically `young', as these molecules are depleted quickly from the gas phase at high densities \citep{hirota_2009, Worthen_2021, suzuki_1992, mcelroy_2013, friesen_2013}. In warm regions around young stellar objects (YSOs), carbon chains can form in the gas phase where dust grains warm sufficiently for methane (CH$_4$) to sublimate from the grains and react with C$^+$ \citep{sakai_2013}. Relative abundances of carbon chains may depend significantly on the initial C/O ratio \citep{seo_2018}, making time-scale arguments based solely on chemistry difficult. Toward some dense cores, differences in the spatial distribution of carbon chains and other carbon-bearing species furthermore indicate that illumination by the interstellar radiation field (IRSF) can produce higher carbon-chain abundances via photochemistry at core edges \citep{spezzano_2016, spezzano_2020}. While the origin of enhanced abundance of carbon chains in cold regions can therefore be attributed to several potential causes, comparison of their distribution and kinematics with those of tracers like \amm\ that remain abundant at high densities can provide insights into the formation and evolution of dense cores and filaments. 

The Taurus Molecular Cloud (TMC) complex is a dark cloud system that has been extensively studied as one of the nearest low-mass star-forming regions, with distances ranging from $d \sim 129$~pc to $d \sim 160$\,pc \citep{Elias_1978, Galli_2018, Galli_2019}. Within the TMC, TMC-1 is a narrow filament of gas of $\sim$0.6\,pc in length and width of $\sim$0.1\,pc \citep{toelle_1981} at $d \sim 140$\,pc \citep{Galli_2019}. 
While several IRAS sources are found near TMC-1, no YSOs are present within the filament \citep{Nutter_2008}.
Many carbon-bearing molecular species have been identified toward TMC-1 \citep{sume_1975, Winnewisser_1979, freeman_1983, navarro_2021, Cernicharo_2021a}, including cyanopolyynes up to $n = 11$ \citep{loomis_2021}. 
Chemical models show that the abundances of carbon-bearing species are consistent with time-scales of only $t \sim$10$^5$\,years \citep{pratap_1997}. 
There is a clear projected offset of $\sim$0.3\,pc in location between emission peaks of carbon chains and tracers like \amm\ along the filament \citep{little_1979}, which has been interpreted as due to an age gradient along the filament \citep{hirahara_1992,suzuki_1992}. 
This makes TMC-1 an ideal candidate to investigate the kinematic properties of a molecular cloud filament that may soon form stars.

In this work, we use the Regularized Optimization for Hyper-Spectral Analysis ({\tt ROHSA}) Gaussian decomposition algorithm to robustly identify three distinct velocity-coherent components in TMC-1 through \hcfn\ (9-8) line emission. We argue that these components trace different layers of the filament, each with their own kinematic behaviour. We propose a model of TMC-1 where the outer layer is inflowing toward the inner layer due to self-gravity, and we find evidence of small-scale fragmentation in the inner layer along the length of TMC-1.

The paper is organized as follows: In Section \ref{sec:data}, we describe the \hcfn\ (9-8) emission and \amm (1, 1) emission observed by the Green Bank Telescope as well as the \hh\ column density maps derived from {\it Herschel} continuum. In Section \ref{sec:rohsa}, we use {\tt ROHSA} to find strong evidence for three kinematically coherent structures in \hcfn\ emission in TMC-1. 
In Section \ref{sec:results}, we investigate the gas properties and in Section \ref{sec:discussion} we conclude that gas in the outer regions of TMC-1 is flowing inwards due to the influence of gravity. Additionally, one of the velocity-coherent structures shows signs of fragmentation that are not visible in the dust continuum. We summarise our findings in Section \ref{sec:conclusion}.

\begin{figure*}
    \centering
    \includegraphics[width=\linewidth]{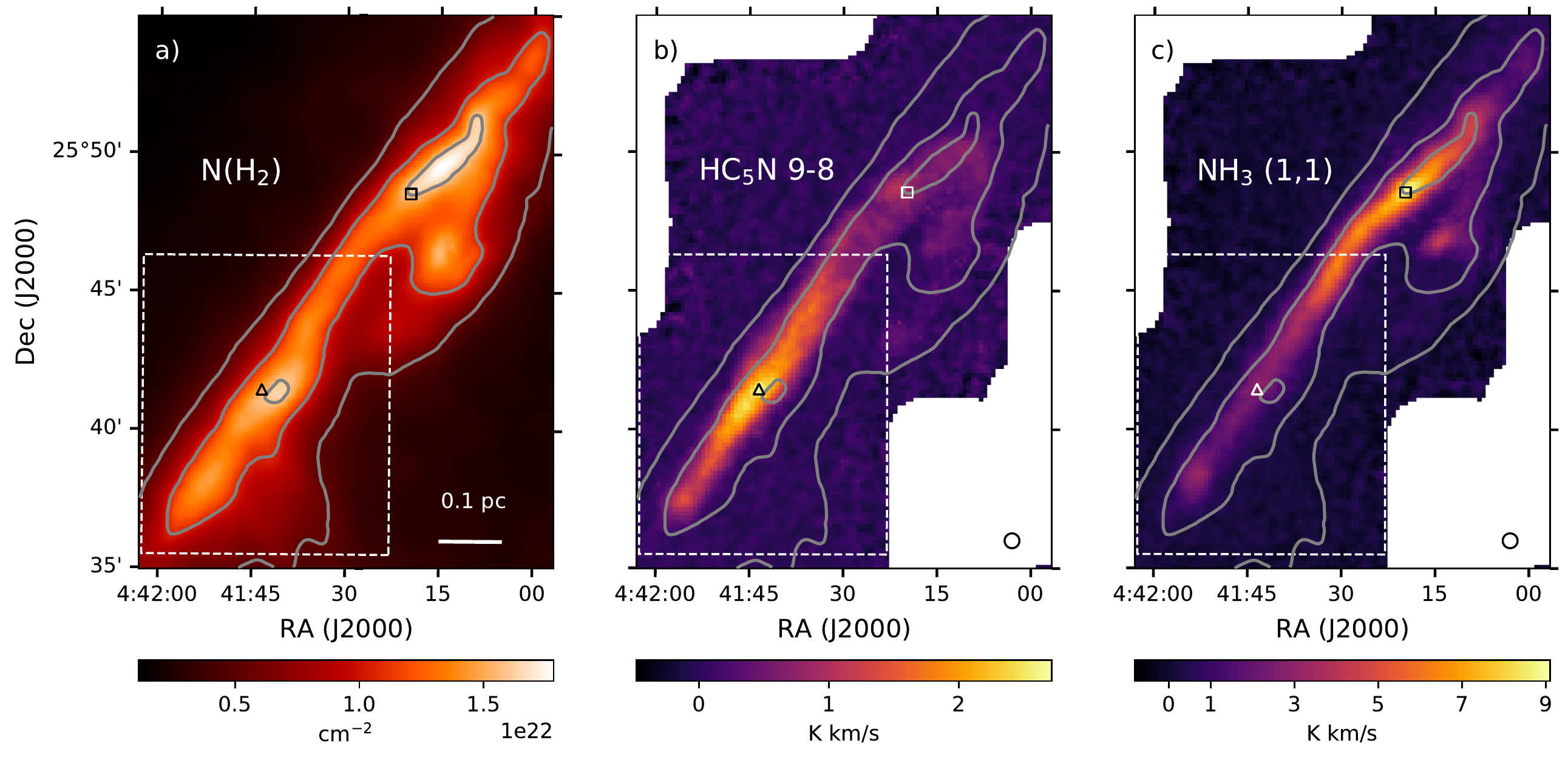}
    \caption{a) $N($H$_2)$ towards TMC-1 (A. Singh et al., in prep.). In all plots, contours are at 0.5, 1.1, and 1.6\,$\times 10^{22}$\,cm$^{-2}$. The peak locations of \hcfn\ and \amm\ from GAS, updated from historical values, are shown by the triangle and square markers respectively. The white dashed box shows the area of TMC-1 discussed primarily in this paper. b) Integrated HC$_5$N 9-8 emission over the same region as in a). The black circle at lower right shows the $\sim 32$\,\arcsec\ GBT beam at 23\,GHz. c) Integrated NH$_3$ (1,1) intensity over the same region as in a) and b), with GBT beam shown at lower right as in b).}
    \label{fig:all_three} 
\end{figure*}

\section{Data}
\label{sec:data}

\subsection{Green Bank Ammonia Survey}
\label{sec:gas}

The Green Bank Ammonia Survey \citep[GAS;][]{Friesen_2017} performed molecular emission mapping of all Gould Belt star-forming regions in the northern hemisphere with $A_V \gtrsim 7$\,mag in \amm\ along with C$_2$S, \hcfn, and HC$_7$N. Here, we analyse  HC$_5$N $J = 9-8$ emission (rest frequency: 23963.9010\,MHz) and emission from the NH$_3$ ($J$, $K$) = (1, 1) inversion transition (rest frequency: 23694.4955\,MHz) towards TMC-1. Fig. \ref{fig:all_three} shows the entire TMC-1 filament in $N(\mbox{H}_2)$, \hcfn\ integrated intensity, and \amm\ integrated intensity. In this paper, we examine primarily the region contained in the dashed white box as it contains strong \hcfn\ emission in a long ($\sim$0.5\,pc) narrow filament, including the historical cyanopolyyne peak \citep[CP;][]{little_1979}. This region excludes the \amm\ peak (identified by the square marker) further to the north-west along TMC-1. At $d = 140$\,pc and 23\,GHz, the $\sim$32\,\arcsec\ (FWHM) GBT beam subtends a physical scale of $\sim$0.02\,pc. The velocity channels have a resolution of 71.58\,m\,s$^{-1}$, and the root mean square (rms) noise is $\sim$0.16\,K per spectral channel. The zeroth-moment (integrated intensity) map is shown in the middle panel of Fig. \ref{fig:all_three} and we calculated the background level, $\sigma_{\text{bg}}$, in this map by taking the standard deviation of a region without any significant emission. This value was determined to be $\sigma_{\text{bg}} = 0.08$\,K\kms.
Full details of the survey including observing details, telescope calibration, data reduction and imaging can be found in Section 2 of the first Data Release (DR1) from the GAS Collaboration \citep{Friesen_2017}.

\subsection{\nh\ from {\it Herschel} dust continuum}
\label{sec:herschel}

The \hh\ column density maps [\nh] were derived by fitting a spectral energy distribution (SED) to  continuum emission from cold dust which was initially mapped by the \textit{Herschel Space Observatory} at 160,  250,  350  and  500\,$\mu$m.
This method was designed to optimize model uncertainties and improve robustness upon previous \nh\ derivations \citep[][A. Singh \& P.G. Martin, in prep.]{singh_2021}. The column density maps have a resolution of 36\,\arcsec, matching well the $\sim$32\,\arcsec\ resolution of the GBT data. We regrid the \nh\ maps to the same pixel scale as the GAS data.

As seen in Fig. \ref{fig:all_three} a), the H$_2$ column density map peaks at 1.7\,$\times$\,10$^{22}$\,cm$^{-2}$ in the centre of the region. 
We calculate the background column density by averaging over four regions of the map that had little to no structured emission and find \nbg\,$= 1.06 \times 10^{21}$\,cm$^{-2}$. Fig. \ref{fig:all_three} shows that the north-east edge of TMC-1 runs fairly straight along the length of the filament while the south-west edge flares outward.

\section{Line Fitting}
\label{sec:linefit}

\begin{figure*}
    \centering
    \includegraphics[width=\linewidth]{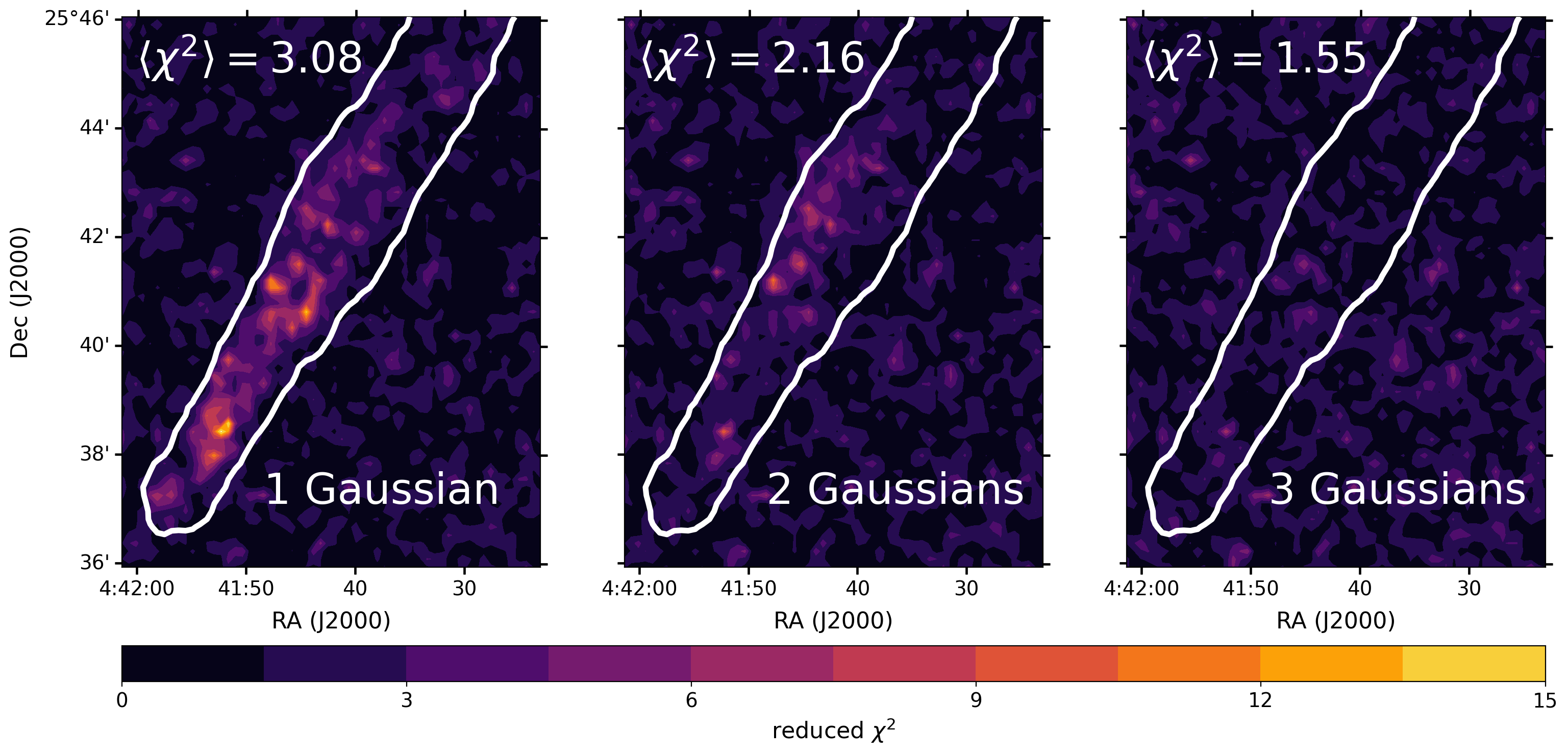}
    \caption{Reduced $\chi^2$ maps for 1-, 2-, and 3-Gaussian fits. The white contour in all plots outlines the region with total integrated intensity $> 5$-$\sigma_{\text{bg}}$ and the average reduced $\chi^2$ value in within the contour is indicated. {\it Left:} Increased $\chi^2$ towards the southern half of the filament indicates that a 1-Gaussian fit is not sufficient to describe the emission. {\it Centre:} Increased $\chi^2$ along the north-eastern edge of the filament indicates that an additional component is necessary. {\it Right:} Low $\chi^2$ within the contour shows that a 3-Gaussian model is a good description of the emission.}
    \label{fig:chi2_1g}
\end{figure*}
\begin{figure*}
    \centering
    \includegraphics[width=\linewidth]{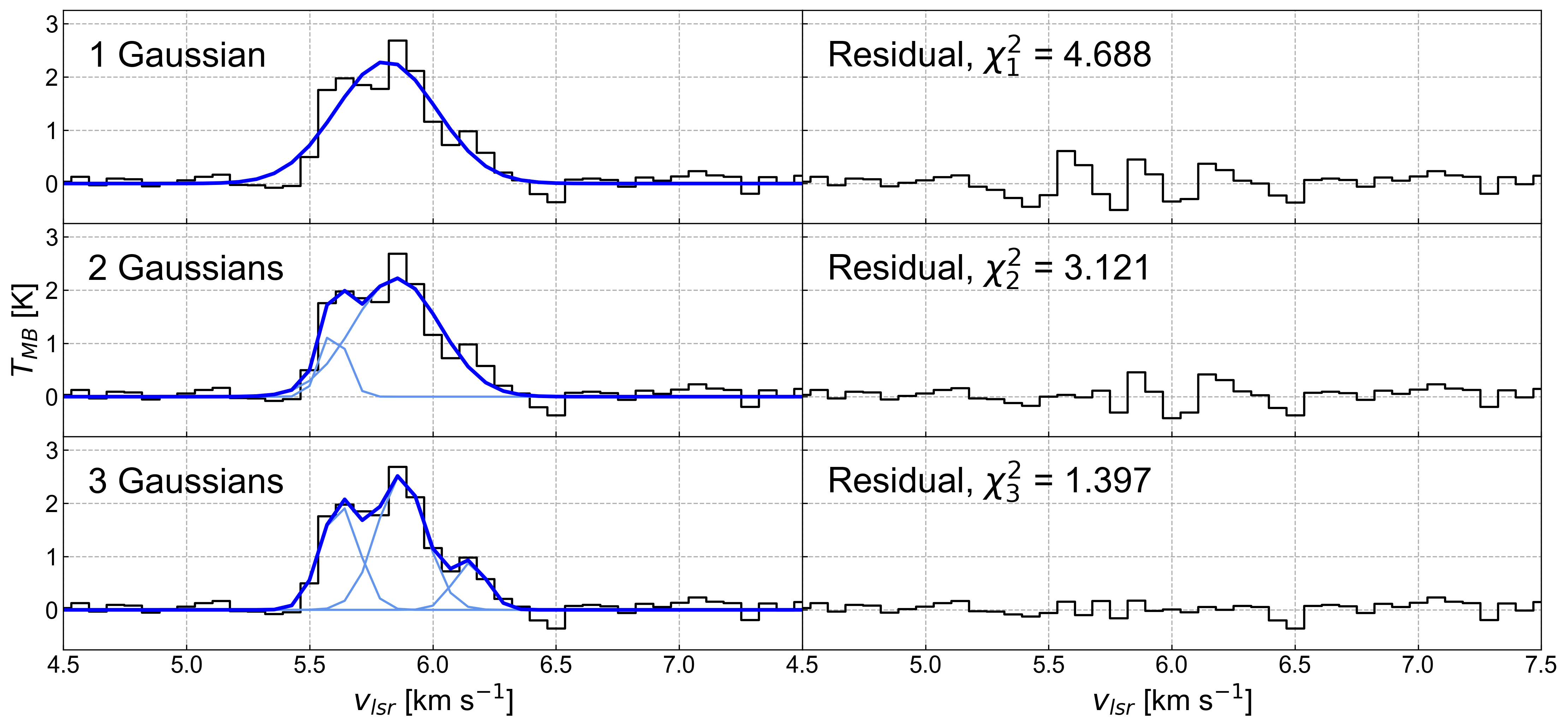}
    \caption{Example of 1-, 2-, and 3-Gaussian fit for a randomly selected pixel. The observations are in black, the multi-Gaussian fit is in dark blue, and the individual Gaussian components are in light blue. {\it Top Row}: Single Gaussian model and residual of the fit with a reduced $\chi^2$ of 4.688. {\it Middle Row}: Two Gaussian model and residual of the fit with a reduced $\chi^2$ of 3.121. {\it Bottom Row}: Three Gaussian model and residual of the fit with a reduced $\chi^2$ of 1.397.}
    \label{fig:spectra}
\end{figure*}

\subsection{Single Gaussian fitting}

We initially fit the \hcfn\ PPV data cube with a single Gaussian component
and use the reduced $\chi^2$ statistic to evaluate the fit in each pixel across the emission map. The leftmost panel of Fig. \ref{fig:chi2_1g} shows the value of the reduced $\chi^2$ statistic across the area of the emission map. The white contour outlines all pixels with a total integrated intensity $> 5$$\sigma_{\text{bg}}$ and within this contour there is an average reduced $\chi^2 = 3.08$. Furthermore, as can be seen in Fig. \ref{fig:chi2_1g}, nearly 24\% of pixels have a reduced $\chi^2 > 4$, signifying that a single Gaussian model is a poor fit to the \hcfn\ emission, particularly in the lower half of the filament. As previously noted, multiple components or emission peaks in carbon-bearing species have been identified towards this region \citep[e.g.,][]{feher_2016, Dobashi_2019}, so it is unsurprising that a single Gaussian does not provide a good fit across TMC-1. We thus expand our fitting to allow for multiple Gaussian components and use a more complex fitting procedure, described in the next section.  

\subsection{{\tt ROHSA}}
\label{sec:rohsa}

{\tt ROHSA} is a Gaussian decomposition algorithm that performs a regression analysis using a regularized non-linear least-square criterion to ensure spatial coherency of the solution \citep[][]{Marchal_2019}.
The algorithm fits the entire PPV cube at once for the desired number of Gaussian components $N$. 
Initialization of the Gaussian parameter maps $\big(\ab_n, \mub_n, \sigmab_n\big)$ is obtained via an iterative multiresolution procedure from a coarse to a fine grid \citep[See section 2.4.3 in][]{Marchal_2019}.
Here, $\ab_{n} \geq \bm{0}$ is the amplitude, $\mub_{n}$ the position, and $\sigmab_{n}$ the standard deviation 2D maps of the $n$-th Gaussian profile across the plane of the sky.
Note that {\tt ROHSA} uses $\mu$ to denote the position or line-of-sight velocity but we will change notation and use \vlsr\ for the remainder of this work.
The minimized cost function includes Laplacian filtering of each parameter map that penalizes small-scale spatial fluctuations. The hyper-parameters $\lambda_{\ab}$, $\lambda_{\mub}$, and $\lambda_{\sigmab}$ tune the balance between the terms. The impacts of varying these hyper-parameters will be explored in Section \ref{sec:stability} through an analysis of the solution consistency.
To ensure spatial continuity of the solution over the field, missing data (due to the finite map extent) at the north-east corner of the \hcfn\ data cube were filled, for each spectrum, with a Gaussian random noise with the same dispersion as the noise in the original data.

We ran {\tt ROHSA} with $\lambda_{\ab}$= $\lambda_{\mub}$=$\lambda_{\sigmab}$=100 and $N = 2-5$ Gaussian components.
This specific set of hyper-parameters was chosen to be high enough to obtain spatially smooth parameter maps but low enough to not alter the native resolution of the observation.
We found that the solution for $N = 3$ and $N = 4$ had similarly good fits to the data with an average reduced $\chi^2$ within the $5$$\sigma_{\text{bg}}$ contour (from Fig. \ref{fig:chi2_1g}) of $\chi^2 = 1.55$ and 1.61, respectively. With their $\chi^2$ values being similar, the simpler model with 3 components is the preferred interpretation. We therefore use a 3-Gaussian model to describe the velocity substructures in the \hcfn\ emission. Fig. \ref{fig:spectra} shows the spectra, the residuals, and the reduced $\chi^2$ statistics for a randomly selected example pixel given a 1-, 2-, and 3-component Gaussian model, demonstrating the detail that {\tt ROHSA} is able to extract. Table \ref{tab:Ng_chi2} in the appendix lists the average reduced $\chi^2$ statistic for models with $N=1-5$. 

\subsection{Velocity Coherent Components}
\label{sec:components}

Unlike other multicomponent Gaussian search algorithms that are able to select the number of components on a per-pixel basis \citep[e.g.,][]{Hacar_2013, MChen_2020}, {\tt ROHSA} simultaneously fits the entire PPV cube, and thus requires that parameters for $N$ Gaussians are found in every pixel. In each component, we say that a given pixel is significant if its integrated intensity is $> 3$$\sigma_{\text{bg}}$.

\begin{figure*}
    \centering
    \captionsetup{justification=centering}
    \begin{subfigure}[b]{\linewidth}
        \includegraphics[width=0.9\linewidth]{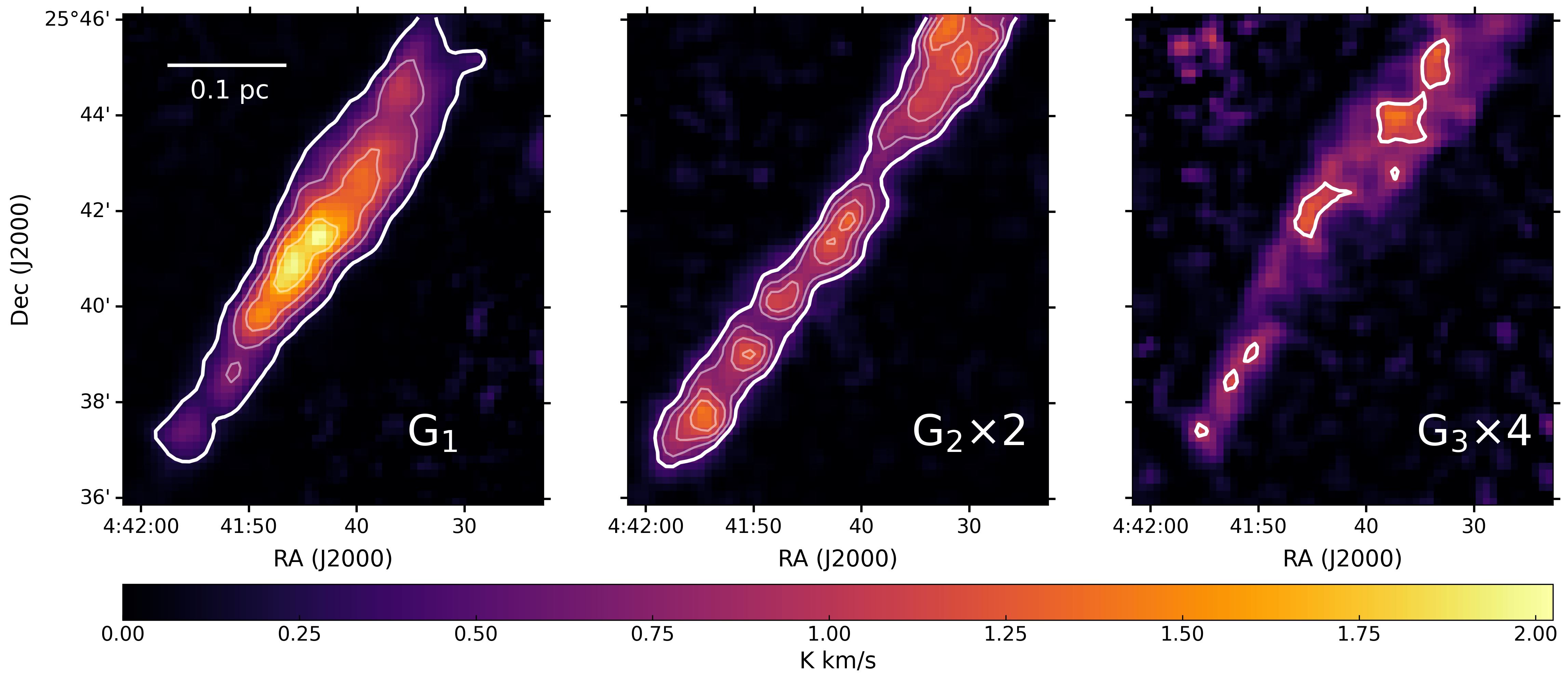}
    \end{subfigure}
    \begin{subfigure}[b]{\linewidth}
        \includegraphics[width=0.9\linewidth]{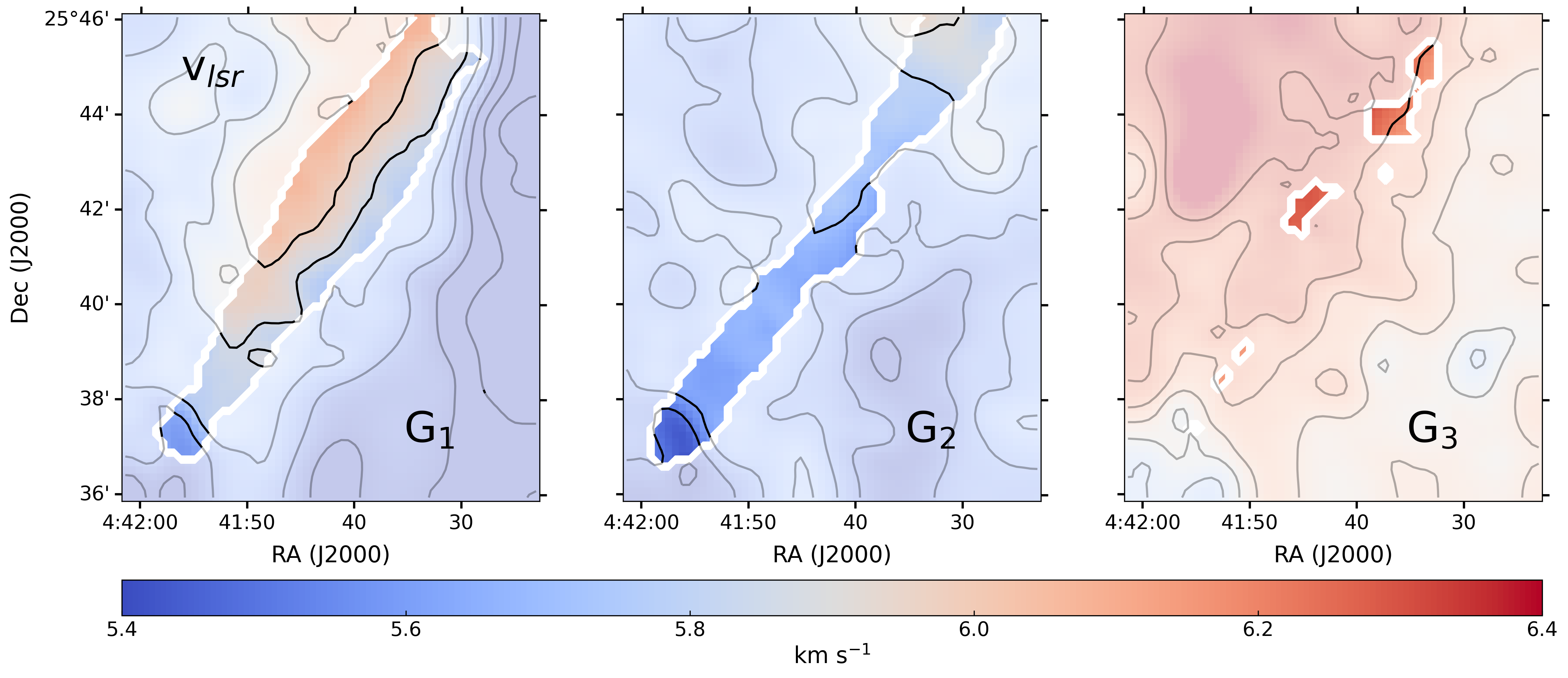}
    \end{subfigure}
    \begin{subfigure}[b]{\linewidth}
        \includegraphics[width=0.9\linewidth]{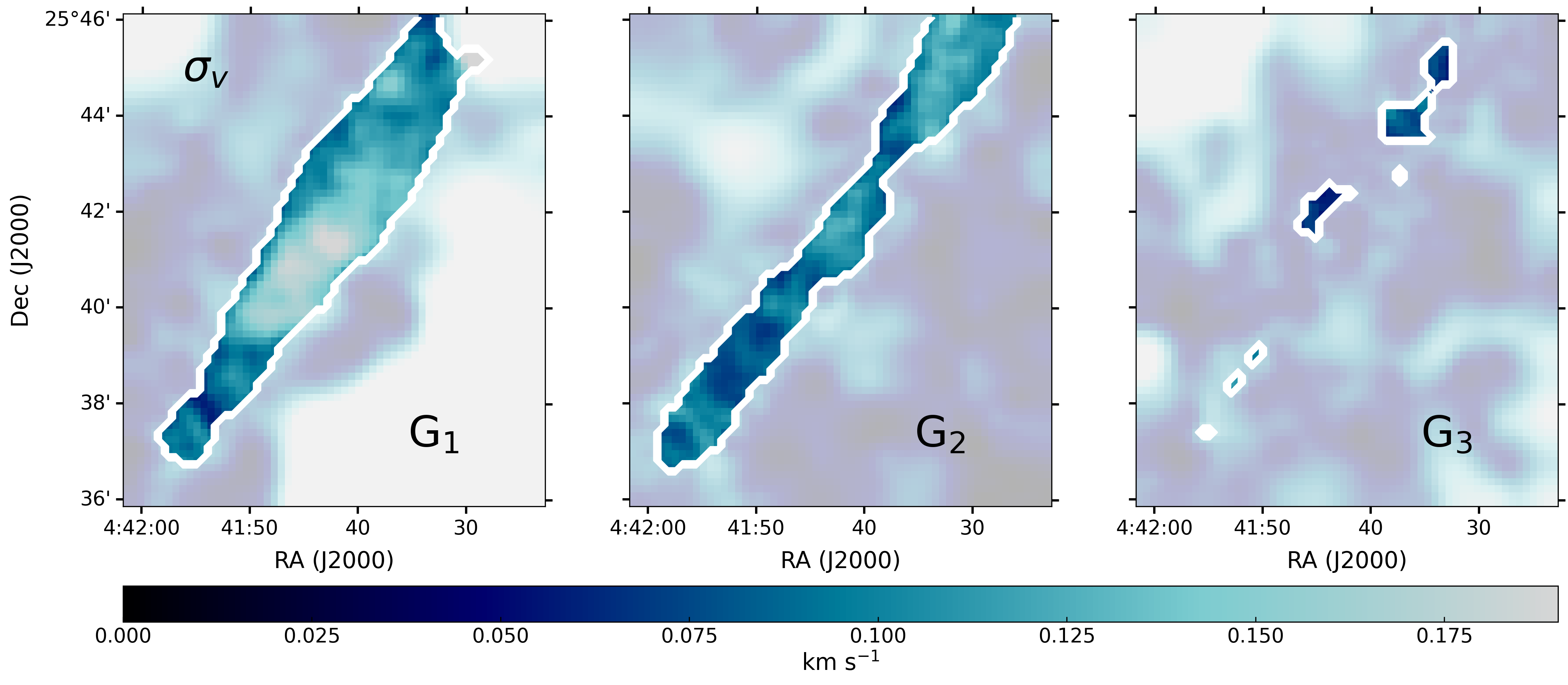}
    \end{subfigure}
    \caption{{\it Top}: Integrated intensity maps for \Gi, \Gii, and \Giii\ where \Gii\ has been multiplied by 2 and \Giii\ has been multiplied by a factor of 4 to accentuate faint details. A 0.1\,pc scale bar is included in the leftmost panel. All components are outlined by a thick white contour outlining the $3$-$\sigma_{\text{bg}}$ boundary. Lighter contours are as follows: \Gi\ at 9, 15, and $21$-$\sigma_{\text{bg}}$, \Gii\ at 4.5, 6, and $7.5$-$\sigma_{\text{bg}}$, and \Giii\ has no contours $\geq4.5$-$\sigma_{\text{bg}}$. {\it Middle}: \vlsr\ maps for \Gi, \Gii, and \Giii. Black contours trace boundaries of equal \vlsr. Those contours are still shown outside of the $3$-$\sigma_{\text{bg}}$ boundary although the velocity fit in those regions is not to be trusted. {\tt ROHSA} fits the entire spectral cube together, so noise is being fitted in regions with no emission. To demonstrate this, We have muted the colours in the exterior regions. {\it Bottom}: Velocity dispersion maps for \Gi, \Gii, and \Giii. Pixels exterior to the $3$-$\sigma_{\text{bg}}$ boundary are similarly muted as in the middle panels.}
    \label{fig:param_maps}
\end{figure*}

Here, we present the general properties of the 3 component {\tt ROHSA} solution with respect to each parameter map, all of which are shown in Fig. \ref{fig:param_maps}. 
We construct a model PPV cube by summing, at each pixel, the three individual Gaussian components fit to the observed spectra by {\tt ROHSA}. Section \ref{sec:stability} analyzes the consistency of the fit results with respect to the noise properties of the data as well as the hyper-parameters that are inputs to {\tt ROHSA}.
In Section \ref{sec:velgrad}, we give a quantitative analysis of the velocity gradients and Table \ref{tab:colden} compiles the mean \vlsr\ and mean velocity dispersion along with column densities (which are calculated in Section \ref{sec:colden}).

\begin{itemize}
    \item[i)] \Gi\ is the brightest component and its $3$$\sigma_{\text{bg}}$ contour traces a very similar shape to the original integrated intensity map. Its integrated intensity peak is located about halfway along the filament with a mean value of 0.79\,K~\kms. It has a mean $v_\mathrm{LSR} = 5.94$\,\kms, with a clear transverse velocity gradient that descends from the north-east down to the south-west edge in the upper two-thirds of the component. The velocity dispersion of \Gi\ is the highest of the three components, with an average value of 0.12\,\kms, peaking around 0.23\,\kms\ in roughly the same position as the intensity peak. The dispersion varies substantially across \Gi\ with a standard deviation of 0.03\,\kms.
    \vspace{1em}
    \item[ii)] The second brightest component, \Gii, is 50\% as bright as \Gi\ on average, with a mean integrated intensity of 0.43\,K~\kms. The morphology of \Gii\ is also relatively straight and narrow,  but shows regularly spaced peaks along the length of the component. The mean $v_\mathrm{LSR} = 5.70$\,\kms\ and has a slight gradient along the filament that will be discussed in more detail in Section \ref{sec:velgrad}. The mean velocity dispersion is $\sim$ 0.1\,\kms\ with heightened velocity dispersion near some of the emission peaks.
    \vspace{1em}
    \item[iii)] The faintest component, \Giii\ is 50\% as bright as \Gii\ on average, and consequently $\sim$25\% as bright as \Gi. \Giii\ emission is found primarily in the upper half of TMC-1, and its $3$$\sigma_{\text{bg}}$ contour has a clumpy and more irregular shape than the two brighter components. This component has a mean $v_\mathrm{LSR} = 6.21$\,\kms\ and an average velocity dispersion of 0.08\,\kms\ which is slightly larger than the spectral resolution (0.072\,\kms).
\end{itemize}

\subsection{Consistency of the {\tt ROHSA} Solution}
\label{sec:stability}

{\tt ROHSA} converges on a solution through its coarse-to-fine grid procedure. Following the methodology developed in \citet{marchal_2021}, we test the consistency of the solution using two independent methods. The goal of this analysis is to examine how the kinematic properties for each of the three Gaussian components would be affected due to changes in either the noise properties of the data or the tuning of the hyper-parameters in the algorithm.
For the first test, we perform 100 runs of the decomposition where we inject Gaussian noise with the same variance as the original data cube into the model data cube. For the second test, we perform 125 runs on the original data cube but systematically varied the hyper-parameters by up to 40\% on either side of their original value. 
Solutions are cross-matched with the original solution by assigning the \Gi, \Gii, and \Giii\ labels to components in order of descending average brightness.
For both catalogues of solutions, we estimate uncertainties by taking the standard deviation of the Gaussian parameters for each component. The total uncertainties are then found by adding the contributions from both the noise injections and the hyper-parameter variations in quadrature. Maps of the uncertainty in each parameter can be seen in Fig. \ref{fig:uncert} and average uncertainties are tabulated (Table \ref{tab:uncerts}) in the appendix.

We use this uncertainty to assess how robustly and consistently {\tt ROHSA} is able to separate the different velocity components. 
We use the velocity uncertainty maps to make this assessment because the aim of this work to identify velocity-coherent structures.
The average uncertainty in the measured \vlsr\ across the \Gi\ map is 0.12\,\kms. Most of the pixels with high uncertainty are contained in a small region in the upper portion of the filament. In the lower two-thirds of the filament, the \vlsr\ uncertainty is below 0.07\,\kms.
The average uncertainty in \vlsr\ for \Gii\ is 0.07\,\kms, indicating that {\tt ROHSA} is extremely consistent when assigning which emission peaks belong to the second velocity component. The uncertainty in \vlsr\ is low for \Giii\ along its lower two-thirds, but has a larger uncertainty on average at 0.25\,\kms. Overall, we find that {\tt ROHSA} is robustly identifying \Gi\ and \Gii\ but is less consistent when identifying \Giii.

\section{Results}
\label{sec:results}

\subsection{\hcfn\ Column Densities}
\label{sec:colden}

We calculated column densities for all three Gaussian components identified by {\tt ROHSA} following the not optically thin approximation in \citet{Mangum_2015}. The total \hcfn\ column density $N$ is then given by

\begin{eqnarray}
    N &=& \frac{3h}{8\pi^3S_{ij}\mu^2}\frac{Q_{\text{rot}}}{g_Jg_Kg_I}\exp\bigg({\frac{E_{u}}{T_{\text{ex}}}}\bigg) \\\nonumber
    &\times& \Bigg[\exp\bigg({\frac{h\nu}{k_BT_{\text{ex}}}}\bigg) - 1\Bigg]^{-1} \int \tau_{\nu}d\nu,
\end{eqnarray}
where S$_{ij}$ = 0.47  is the relative line strength and $\mu$ = 4.33\,debye is the dipole moment \citep{Muller_2005}. The rotational partition function $Q_{\text{rot}}$ is computed for the first 100 states in a rigid rotor approximation where the energy of a given state $E_j = hB_0 J(J+1)$, with $B_0$ = 1331.3327\,MHz \citep{Muller_2005}. $Q_{rot}$ is able to be calculated because we make assumptions for \tex\ which is the only free parameter in this equation. The degeneracy functions for a linear molecule are $g_J = 2J_{u}+1, g_K = 1, g_I = 1$. By assuming a Gaussian profile for the line emission, the integral over the optical depth $\tau_{\nu}$ is represented in the final term as

\begin{equation}
    \int \tau_{\nu}d\nu = \sqrt{2\pi}\sigma_v \tau,
\end{equation}
where $\sigma_v$ is the velocity dispersion. The optical depth, $\tau_{\nu}$, is calculated on a per-pixel basis when deriving the column density maps using

\begin{equation}
    \tau_{\nu} = -\ln \Bigg[1 - \frac{T_{mb}}{J_{\nu}(T_{ex}, \nu) - J_{\nu}(T_{bg}, \nu)}\Bigg], \label{eqn:tau}
\end{equation}
where $J_{\nu}$ is the Rayleigh-Jeans equivalent temperature given by $J_{\nu}(T, \nu) = h\nu / k\Big[e^{\frac{h\nu}{kT}} - 1\Big]^{-1}$, $\nu$ is the line rest frequency, $T_{\text{mb}}$ is the line amplitude, \tex\ is the excitation temperature, and $T_{\text{bg}}$ = 2.73\,K is the background temperature due to the cosmic microwave background. Both $Q_{\text{rot}}$ and $\tau_{\nu}$ are dependent on \tex, but as we are only measuring one transition of \hcfn, we are unable to calculate \tex\ directly. Fortunately, we are able to estimate a reasonable range for \tex\ given the nature of the environment. 

The kinetic gas temperature ($T_{\text{kin}}$) across TMC-1 is derived from the \amm\ data following \citet[][]{Friesen_2017}. 
We find a mean $T_{\text{kin}} = 9.6 \pm 1.2$\,K across TMC-1. We take the mean value to be the upper bound on \tex\ for the \hcfn\ line. Additionally, the solution to equation (\ref{eqn:tau}) becomes unphysical toward the brightest \hcfn\ emission for $T_{\text{ex}} < 7.2$\,K, and we thus take this to be the lower bound on the possible value for \tex. We average the upper and lower bounds and take $T_{\text{ex}} = 8.4$ $\pm$ 1.2\,K for \hcfn\ column density calculations and it's uncertainties. Encouragingly, \citet[][]{Bell1998} found the excitation temperature for a similar cyanopolyne chain, HC$_9$N, to be in a similar range, from 7.5 to 8.7\,K.

We calculated optical depth maps for each component across TMC-1 using \tex\ in the range between 7.2\, and 9.6\,K at intervals of 0.1\,K. At lower values of \tex, the opacity increases. \Gi\ is the primary contributor to the overall optical depth, but remains largely optically thin even with low \tex, where the average opacity ranges from 0.4 at \tex\ = 9.6\,K up to 0.8 at \tex\ = 7.2\,K. The average opacity in \Gii\ never exceeds 0.4, and the average opacity in \Giii\ never exceeds 0.2. Additionally, the maximum opacity in \Gi\ ranges from 1.0 (with $T_{\text{ex}} = 9.6$\,K) up to 4.9 (with $T_{\text{ex}} = 7.2$\,K), but these high opacities are found solely toward a compact region around the CP.  We also computed the opacities for a single Gaussian fit in order to estimate the total optical depth along each line of sight. Here, the average opacity ranges from 0.5 to 1.0 and the highest regions of opacity continue to be restricted to a compact region near the CP. The generally low opacity indicates that the multiple velocity components identified by {\tt ROHSA} across the entire filament are real rather than being artefacts of self-absorption. The peak and on-filament mean values for the column density in each component are presented in Table \ref{tab:colden} along with the mean velocities and mean velocity dispersions.

\begin{table*}
    \centering
    \begin{tabular}{c|c|c|c|c|c|c}
        \hline
        Property: & Peak & On-Fil Mean & $\langle$\vlsr$\rangle$ & $\langle$$\sigma{_v}$$\rangle$ & Offset & FWHM \\
        (Unit) & (10$^{12}$ cm$^{-2}$) & (10$^{12}$ cm$^{-2}$) & (\kms) & (\kms) & (\arcsec) & (\arcsec) \\
        \hline \hline
        \Gi & 25$^{+3.4}_{-0.4}$ & 7.4$^{+2.4}_{-0.7}$ & 5.94 $\pm$ 0.09 & 0.12 $\pm$ 0.03 & 14 $\pm$ 2 & 32 $\pm$ 3\\
        \Gii & 6.2$^{+1.1}_{-0.4}$ & 3.3$^{+0.7}_{-0.1}$ & 5.70 $\pm$ 0.11 & 0.10 $\pm$ 0.02 & 10 $\pm$ 3 & 28 $\pm$ 1\\
        \Giii & 2.7$^{+0.3}_{-0.1}$ & 2.1$^{+0.1}_{-0.1}$ & 6.21 $\pm$ 0.07 & 0.08 $\pm$ 0.02 & -- & -- \\
        \hline
    \end{tabular}
    \caption{Properties of the three gaussian components selected by {\tt ROHSA}, \Gi, \Gii, and \Giii. The column density maps were calculated using \tex\ = 8.4\,K and the uncertainties reflect the results of the calculations using the lower and upper bounds on \tex, 7.2\,K and 9.6\,K respectively. 
    We also present the mean \vlsr\ and the mean velocity dispersion values in \kms\ where uncertainties are the standard deviations of the given parameter map. Additionally, offset (from the \nh\ profile centroid) and FWHM measurements are taken from Gaussian fitting to the transverse column density profiles. Uncertainties are found by performing this fitting for map rotations in the range from $-$39\textdegree\ to $-$33\textdegree\ at intervals of 1\textdegree\ and taking the standard deviation of each property.}
    \label{tab:colden}
\end{table*}

\begin{figure*}
    \centering
    \includegraphics[width=\linewidth]{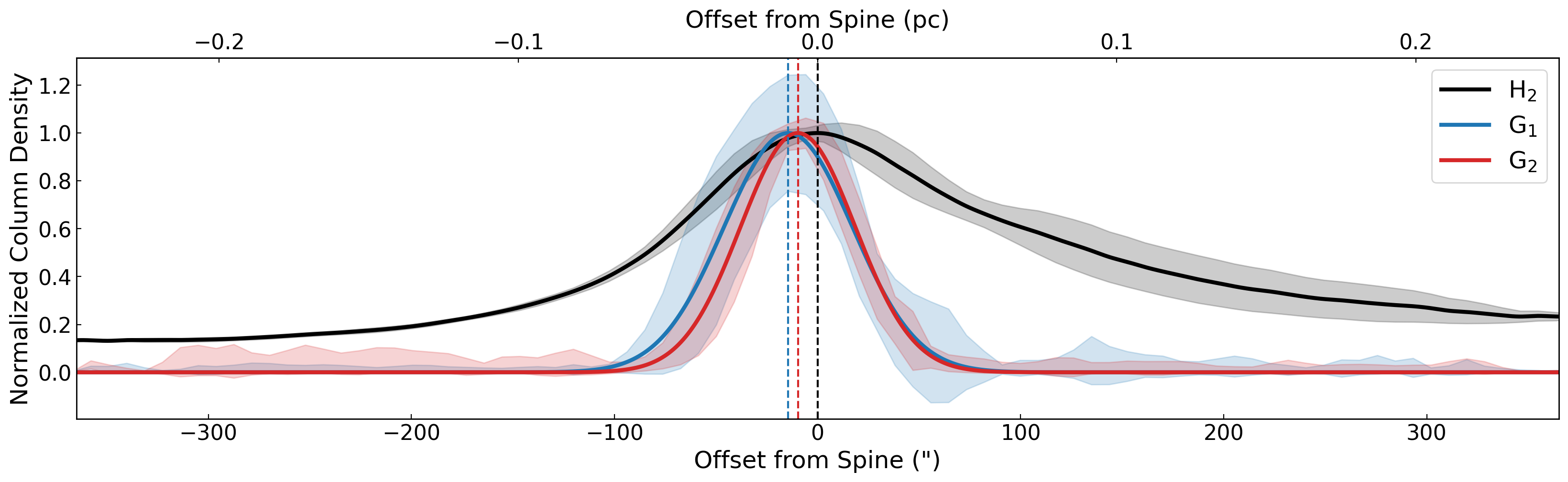}
    \caption{Average normalized transverse column density profiles for \hh\ (black), \Gi\ (blue), and \Gii\ (red). 1-$\sigma$ standard deviations are shown as the shaded regions enclosing each profile curve and vertical dashed lines indicate the profile centroids.}
    \label{fig:all_profiles}
\end{figure*}

In the following sections we focus our analysis on the \Gi\ and \Gii\ components. 
The \Giii\ component is real, as indicated by the reduced $\chi^2$ calculation, and still has emission peaks that are significant ($>3$$\sigma_{\text{bg}}$). However, its regions of significance are clumpy and amorphous (see Fig. \ref{fig:param_maps}) and many of its significant pixels have velocity dispersions smaller than the spectral resolution, making its interpretation uncertain.
In Section \ref{sec:envspin}, we compare the spatial and kinematic properties of \Giii\ to velocity components found in other works. Higher signal-to-noise observations may be necessary to understand the kinematics and faint structure of \Giii.
\Gi\ and \Gii\ contain far more prominent emission and, as we will see, present differences in their distribution and kinematics that will guide the discussion for the remainder of the paper.

\subsection{Spatial Correlation of \hcfn\ and \hh}
\label{sec:align}

We examine the column density profiles, relative positions and orientations of the {\tt ROHSA}-identified \hcfn\ components with respect to \nh. TMC-1 is remarkably linear so we rotate all relevant data maps by $-$36\textdegree\ such that the filament is aligned vertically in our new frame of reference.

We determine the transverse \nh\ profile by normalizing the \nh\ to its peak value along each row of the rotated map. We then find the average profile along the length of the filament and re-scale the profile such that it peaks at unity. The normalized \nh\ profile, shown in Fig. \ref{fig:all_profiles}, is asymmetric, with a steeper drop in \nh\ toward the north-east (in the non-rotated frame) than in the south-west. We therefore do not use a Gaussian profile to determine the filament centre or width. Instead, we first interpolate between data points to identify the profile peak location, which we label the `spine' of the filament. All subsequent positional measurements are made relative to this point. We find the FWHM to be 205 $\pm$ 6\,\arcsec (0.139 $\pm$ 0.004\,pc).

The rotated \Gi\ and \Gii\ column density maps are similarly normalized per-row and averaged along the length of the filament to produce column density profiles. Both components are found preferentially on the north-east (NE) side of the spine. This can be seen in Fig. \ref{fig:all_profiles} where the centre of the x-axis corresponds with the spine of the filament. Both profiles are fitted with Gaussian functions, from which we determine their centroids and FWHM. The widths of the two components are nearly identical, with \Gi\ having a FWHM of 32 $\pm$ 3\arcsec\ (0.022 $\pm$ 0.002\,pc) and \Gii\ having a FWHM of 28 $\pm$ 1\arcsec\ (0.019 $\pm$ 0.001\,pc), 
and are on the same order as widths of filaments found in the Barnard 5 system \citep[$\sim$0.03\,pc;][]{Pineda_2011, Schmiedeke2021}, as traced by \amm\,(1, 1) and (2, 2).
Both components are offset to the NE of the spine with \Gi\ being offset by 14 $\pm$ 2\arcsec\ (0.010 $\pm$ 0.001\,pc) and \Gii\ being offset by 10 $\pm$ 3\arcsec\ (0.007 $\pm$ 0.002\,pc). All uncertainties were estimated by carrying out these calculations for a range of rotation angles between $-$39\textdegree\ and $-$33\textdegree\ and taking the standard deviation of each property. These offsets are small, on the order of a single pixel ($\sim$ 9\arcsec), but they are consistently measured for all rotation angles considered, indicating that \Gi\ and \Gii\ are spatially positioned distinctly on the NE edge of the TMC-1 filament, in contrast to the \nh\ distribution, where the filament is more extended on the south-west side of the spine.

\subsection{Velocity Gradients}
\label{sec:velgrad}

\begin{figure*}
    \centering
    \includegraphics[width=\linewidth]{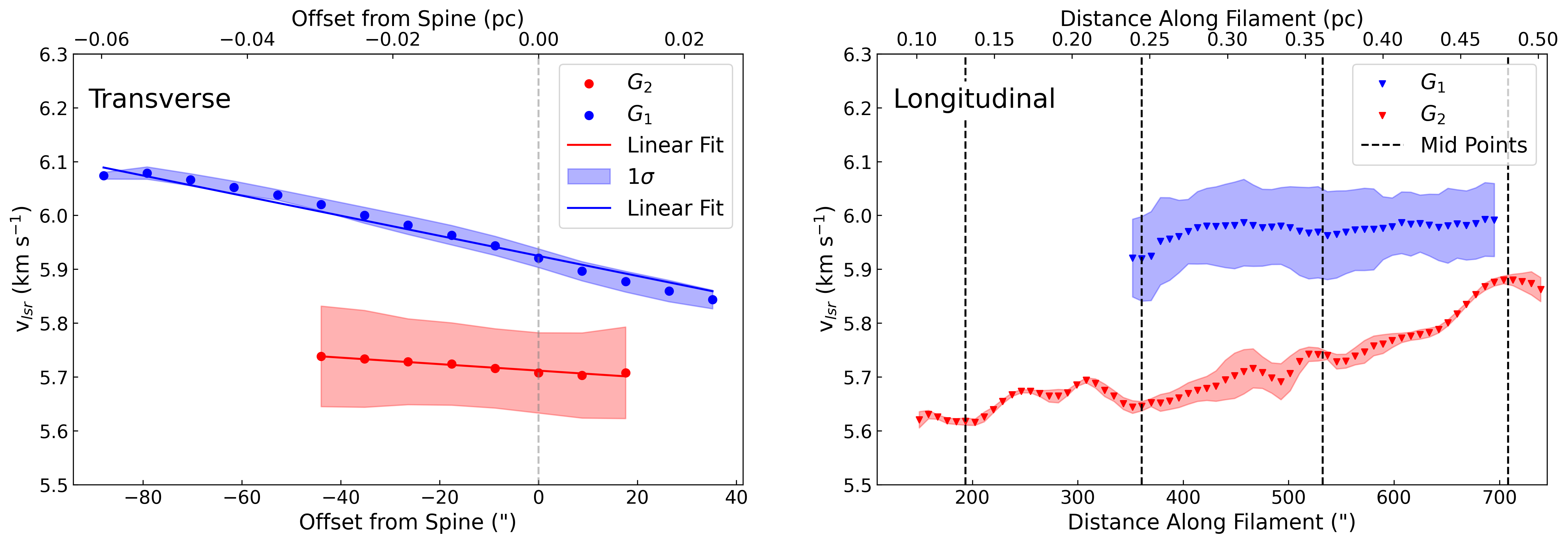}
    \caption{{\it Top}: Transverse Velocity profile for the \Gi\ (blue) and \Gii\ (red) components. The shaded regions show the $1\sigma$ standard deviation of the data and the solid lines are the linear best fits for each component. We also denote the spine's position with a grey, vertical dashed line. {\it Bottom}: Longitudinal velocity profile for the \Gi\ (blue) and \Gii\ (red) components. The shaded regions show the $1\sigma$ standard deviation of the data. The distance along the filament is defined in arcseconds measured from the southern tip of the \hh\ emission. We also show dashed black lines that represent the mid points between pairs of emission peaks in \Gii. They are calculated in Section \ref{sec:clumps} and we posit that they are indicators of density peaks in \nh. We also note that they appear to line up with minima and maxima in the quasi-oscillatory velocity profiles.}
    \label{fig:slopes}
\end{figure*}

As noted in Section \ref{sec:components}, \Gi\ and \Gii\ have significantly different features in \vlsr. Pictured in Fig. \ref{fig:param_maps}, the \Gi\ component has a clear transverse velocity gradient ($\nabla$\vlsr) that extends along a length of 0.25\,pc within the 3$\sigma_{\text{bg}}$ contour. Looking down the filament, the mean \vlsr\ and associated standard deviation is computed for each column, which is then fit with a linear model. To determine the slope of the velocity gradient in \Gi, we only include pixels in the upper two-thirds of the filament which, referring back to the velocity maps in Fig. \ref{fig:param_maps}, is the region containing the strong gradient. The slope of the linear fit, as seen in Fig. \ref{fig:slopes}, is $2.7 \pm 0.1$\,\kmspc, which gives a difference in \vlsr\ of 0.23\,\kms\ across the width of \Gi. 
The average \Gii\ profile is prepared in a similar manner but we use the entire length of the filament as there is no obvious transverse gradient visible in the velocity map. 
The \Gii\ velocity profile has a linear best fit with a slope of $0.8 \pm 0.1$\,\kmspc\ which corresponds to a total change in \vlsr\ of 0.03\,\kms\ across the width of \Gii. Error bars on both measurements come from the linear least-squares fit performed to estimate the slope and intercept parameters. We only consider pixels that fall within the 3$\sigma_{\text{bg}}$ contour which is why the \Gi\ profile extends further on either side of the spine than the \Gii\ profile.

Both gradient measurements fit in well with measured gradient slopes from the literature. \citet{MChen_2020} reports $\nabla$\vlsr\ values $>2$ \kmspc\ across filaments on small scales in NGC 1333 while \citet{fernandez_2014b} reports a typical gradient range of $0.2-2$\,\kmspc\ in Serpens South. On the other hand, typical velocity differences of 0.1$-$0.2\,\kms\ across simulated filaments have been reported in \citet{CChen_2020}.

For the longitudinal velocity gradients we use the southern tip of the filament to be our reference and measure the velocity profiles towards the northern tip (see Fig. \ref{fig:slopes}). Here, the general trends of the gradients are reversed with \Gi\ having a nearly flat profile while the \Gii\ profile has an overall velocity difference of 0.26\,\kms\ along the length of the filament. \Gii\ shows quasi-oscillatory behaviour as well which will be discussed further in Section \ref{sec:clumps}. This type of behaviour has been reported in Taurus L1495/B213 \citep{tafalla_2015} and Serpens South \citep{fernandez_2014b}.

In Fig. \ref{fig:slopes}, there is an interesting relationship between the shaded $1$$\sigma$ standard deviation error bars and the velocity profile along the transverse and longitudinal axes. 
In \Gi, the transverse profile has very small error bars while the longitudinal profile has large error bars, indicating that the coherent trends in velocity are transverse. 
The opposite is true for \Gii, where the longitudinal profile has very small error bars, indicating that the coherent velocity trend is along the the transverse axis. 
We investigate this further in the \Gi\ component by subtracting the transverse linear fit from the \vlsr\ map. We recalculate the $1$$\sigma$ standard deviation on the residual and find it's average (with little variation) to be $\sim$14\,m\,s$^{-1}$, which is much smaller than the velocity resolution of the data ($\sim$72\,m\,s$^{-1}$). This indicates that the dispersion in the \Gi\ longitudinal velocity profile in Fig. \ref{fig:slopes} is driven primarily by the velocity gradient, rather than any small-scale variations in \vlsr.

The inclination angle of the filament on the plane of the sky has a significant effect on the observed velocity gradients. As we are measuring line-of-sight velocities, the existence of any coherent gradient strongly suggests that the filament is not flat in the plane of the sky. This possibility will be explored in more detail in Section \ref{sec:gravcon}.

\section{Discussion}
\label{sec:discussion}

\subsection{Gravitational Stability}
\label{sec:gravcon}

For an isothermal, self-gravitating gas cylinder, the critical mass-per-unit-length \citep[][]{Ostriker_1964} that can be supported by thermal motions is given by

\begin{equation}
    M_{\text{lin,crit}} = \frac{2c_s^2}{G}, \label{eqn:ostriker}
\end{equation}
where $c_s = \sqrt{\frac{k_B T_{\text{kin}}}{\mu m_\text{p}}}$ is the isothermal speed of sound. In this model, the only two parameters that determine the mass-per-unit-length are the temperature $T$ and the mean molecular weight per free particle \citep[$\mu=2.37$,][]{Kauffmann_2008}. As stated in Section \ref{sec:colden}, the average kinetic temperature across TMC-1 (derived from \amm) is $T_{\text{kin}} =9.6 \pm 1.2$\,K. Following Equation \ref{eqn:ostriker} critical line mass is then $M_{\text{L,crit}} = 15.5 \pm 2.0$\,M$_{\odot}$~pc$^{-1}$. 

The motion of the gas is also influenced by non-thermal (potentially turbulent) processes which may contribute additional support to the filamentary structure, adding a term to the expression for the critical line mass:
\begin{equation}
    M_{\text{lin,vir}} = \frac{2(c_\text{s}^2+\langle \sigma_{\text{nt}}\rangle^2)}{G},
\end{equation}
where $\langle \sigma_{\text{nt}}\rangle$ is the average non-thermal velocity dispersion in the \Gi\ component.
The non-thermal dispersion is found by subtracting the thermal dispersion of \hcfn\ from the observed velocity dispersion in quadrature: 

\begin{equation}
    \sigma_{\text{nt}} = \sqrt{\sigma_{v}^2 - \sigma_{t, \mathrm{HC_5N}}^2}.
\end{equation}

\noindent The thermal contribution from \hcfn\ is given as

\begin{equation}
    \sigma_{\text{t,HC$_5$N}} = \sqrt{\frac{k_BT_{\text{kin}}}{\mu_\mathrm{HC_5N}}},
\end{equation}

\noindent where $\mu_\mathrm{HC_5N} = 75.0681$\,amu is the mass of \hcfn\ and $T_{\text{kin}} =9.6$\,K being taken as the average temperature across the entire filament. This increases $M_{\text{lin,vir}}$ to $21.7 \pm 2$\,M$_{\odot}$~pc$^{-1}$. 
Broadening of the line width is not always due to random, turbulent motions, however. If the motions originate from infalling gas, then some portion of the non-thermal motions will not contribute to supporting the filament. The true $M_{\text{lin,vir}}$ likely lies somewhere between $\sim$16~M$_{\odot}$~pc$^{-1}$ and $\sim$22~M$_{\odot}$~pc$^{-1}$ (including thermal only, and thermal and non-thermal contributions, respectively).

Magnetic fields can also provide support against gravitational collapse.
Infrared polarization measurements indicate that the magnetic field in the plane of the sky (pos) is perpendicular to the long axis of TMC-1 \citep{tamura_1987}. \citet{chapman_2011} determine the magnetic field strength $\langle B_{\text{pos}} \rangle = 42 \pm 4~\mu$G toward Heiles Cloud 2 using the Chandrasekhar-Fermi method, but note that the measurement is made using small numbers of polarization vectors. 
Via Zeeman splitting measurements, \citet{troland_2008} measure the line-of-sight (los) component of the magnetic field toward TMC-1, finding $\langle B_{\text{los}} \rangle = 9.1 \pm 2.2~\mu$G, 
showing that the magnetic field is largely in the plane of the sky.
This suggests that TMC-1 and the larger Heiles Cloud 2 formed by contraction along the magnetic field lines \citep{tamura_1987}.
\citet{tomisaka_2014} show that the critical line mass of a filament threaded by a perpendicular magnetic field $B_0$, like TMC-1, is given by $M_{\text{L,B}} \simeq 22.4 (R/0.5~\mathrm{pc})(B_0/10~\mu\mathrm{G})\simeq 11$\,M$_\odot$~pc$^{-1}$. The critical line mass combining thermal, non-thermal, and magnetic contributions is then $M_{\text{L,$\mathrm{crit}$,B}} = M_{\text{L,$\mathrm{crit}$}} + M_{\text{L,B}} 
\sim 27$ -- $33$\,M$_\odot$~pc$^{-1}$.
We note that there are no higher resolution submillimeter polarization measurements toward TMC-1 directly, and the uncertainty in $B_{\text{pos}}$ (and hence $M_{\text{L,B}}$) is likely greater than the measurement uncertainty suggests.



From the continuum data, we estimate the physical line mass, $M_{lin}$, of the filament with

\begin{equation}
    M_{\text{lin}} = \frac{A_{\text{pix}}}{L}\sum^{\text{fil}} N(\text{H$_2$}),
\end{equation}
where $A_{\text{pix}}$ is the physical area of a pixel and $L$ is the length of the filament. Here, we approximate the filament as a cylinder with diameter equal to the FHWM (0.12\,pc, computed in Section \ref{sec:align}) so all pixels within 0.06 pc of the spine are included in the sum.
We find $M_{lin} = 25.5$\,M$_{\odot}$~pc$^{-1}$, making the line mass 1.2$-$1.6$\times$ the critical value without including magnetic field support. Other studies that have shown TMC-1 is trans- or super-critical \citep{feher_2016, Dobashi_2019}. The observed line mass is uncertain due to projection effects, but TMC-1 would have to be inclined at a fairly extreme angle ($i >$ 60\textdegree) to actually be sub-critical.
Including magnetic field support, TMC-1 may be trans-critical. 
Using similar magnetic field strength estimates, \citet{nakamura_2019} argue that the filament is magnetically super-critical and remains unstable to gravitational collapse. 

Based on the above discussion, the filament may be unstable to gravitational collapse. The free-fall time-scale describes how long it would take for the filament to collapse to a line under the sole influence of self-gravity.
For a collapsing filament, \citet{Pon_2012} derived the free-fall time for a collapsing, constant density cylinder, $\tau_{\text{1D}}$ as

\begin{equation}
    \tau_{\text{1D}} = \sqrt{\frac{32A}{\pi^2}}\tau_{\text{3D}},
\end{equation}
where $\tau_{\text{3D}}$ is the free-fall time-scale for a uniform-density sphere with the same density as the filament and $A$ is the aspect ratio ($A = 5$ for TMC-1). The density (needed to find $\tau_{\text{3D}}$) is $\rho = 4\times10^4$ cm$^{-3}$, given the \nh\ profile and assuming a cylindrical model for the filament. This gives a free-fall time of $\tau_{\text{1D}} \approx 0.6$\,Myr.
The corresponding free-fall velocity gradient across the width of the filament is $\sim$2~km~s$^{-1}$, in good agreement with the measured value.
For gravitational contraction, \citet{CChen_2020} show that $\dot{r}^2 = GM(r)/L$, predicting a velocity gradient of $\sim$5.6\,km~s$^{-1}$~pc$^{-1}$ across the width of TMC-1, a factor $\lesssim$2 greater than the measured value.

Velocity gradients across filaments may originate in several different bulk motions, including rotation, shearing, and gravitational inflow in a sheet.
%
%
Studies of bulk-rotation in galactic molecular clouds \citep{phillips_1999, braine_2018} suggest that rotational support is not often significant enough to hinder cloud collapse. We follow a similar methodology as \citet{braine_2018} to investigate the importance of rotation in the TMC-1 filamentary system. The ratio of rotational energy ($\mathcal{E}_{\text{rot}}$) to gravitational energy ($\mathcal{W}$) is:

\begin{equation}
    \frac{\mathcal{E}_{\text{rot}}}{\mathcal{W}} \equiv \frac{\frac{1}{2}I_z \omega^2}{GM_{\text{lin}}^2L}.
\end{equation}
Here, $I_z = \frac{2\pi}{3}\rho R^5$ for a constant density cylinder and the angular velocity $\omega$ = $8.8 \times 10^{-14}$\,rad~s$^{-1}$. We find that the rotational energy is only about 1\% of the gravitational energy and thus argue that any structural support being provided by bulk-rotation is negligible. This does not rule out the possibility that some of the velocity gradient is due to rotation, but the period of rotation would be 2.3\,Myr. This is nearly an order of magnitude longer than the expected free-fall time. Given the dominance of gravitational potential energy over rotational kinetic energy, it is unlikely that rotation drives the transverse velocity gradient.

Another interpretation is shearing motions where the observed velocity gradient would indicate that the structure is expanding and would dissipate on a time-scale of $\approx$0.4\,Myr. However, TMC-1 is embedded in a cold, dense molecular cloud and, as previously stated, is documented to be trans- or super-critical \citep{feher_2016, Dobashi_2019}. We therefore find it unlikely that the velocity gradient is indicative of dissipation.

Since solid-body rotation and shear are unlikely to be responsible for the significant gradient we identify across TMC-1, we next investigate the role of self-gravity. \citet{CO_2014, CO_2015} investigated a two-stage formation channel for filaments wherein small velocity perturbations at the interface of large-scale flows instigates self-gravitating overdensities. These overdensities then pull material inwards from the surrounding medium. Often, these overdensities are planar, and the direction of gas flow is along the plane of the overdensity, resulting in an anisotropic inflow of material and an observable transverse velocity gradient (as long as the viewing angle is not parallel or perpendicular to the plane). However, transverse velocity gradients can also be generated in filaments formed directly from compression of local shocks. 
\citet[][]{CChen_2020} present a dimensionless coefficient, $C_v$, to differentiate between velocity gradients driven by shock compression vs. those resulting from anisotropic gravitational accretion:

\begin{equation}
    C_v \equiv \frac{\Delta v_\text{h}^2}{GM_{\text{lin}}}.
\end{equation}

\noindent Here, $\Delta v_h$ is half the velocity difference across the width of the filament and $M_{\text{lin}}$ is the mass-per-unit-length. This coefficient compares the relative contribution of turbulent motions with the local strength of gravity. If $C_v \gg 1$, the filament is forming as the consequence of shock compression and the local turbulence far exceeds gravitationally-induced kinematics. On the other hand, if $C_v \lesssim  1$, the local gravity is comparable to local turbulence and the motion of the gas is suggestive of gravitational accretion, wherein material is pulled inwards along the plane of the forming filament. For the velocity gradient of \Gi\ and $M_{\text{lin}}$ for TMC-1 determined in Section\ \ref{sec:gravcon}, we calculate a value of $C_v = 0.21 \pm 0.10$, which supports a model where the observed velocity gradient is the result of gravitational-induced gas inflows. 
This coefficient is dependent on the inclination angle, $i$, of the transverse axis with respect to the plane of the sky by a factor of (sin $i$)$^{-2}$. Although we do not know the inclination angle, we find that $i$ must be less than 10\textdegree\ for $C_v$ to be significantly greater than unity.
We therefore argue that gravitationally induced inflow is the main source of the transverse velocity gradient and self-gravity is the primary driver of the evolution of TMC-1.

\subsection{Envelope-Spindle Model}
\label{sec:envspin}

The gravity-driven velocity gradient and trans-critical $M_\text{L}$ imply that the filament is accreting anisotropically at large radii ($r > ~0.06$\,pc). Specifically, Fig. 1 in \citet[][]{CChen_2020} indicates the likely physical picture where the central region of the filament is surrounded by a flattened, outer region of gas that extends on both sides. In this model, the gas in the extended regions are inflowing towards the centre of the filament. Additionally, the off-centre distribution of the \hcfn\ to the NE side of the filament implies that there is an asymmetry present in the environment. It could be kinematically driven, where the accretion of low-density material is only occurring on the NE side of the filament. This would cause \hcfn\ to be preferentially formed where new, carbon-rich material is being accreted onto the filament. It could also be due to uneven illumination with the ISRF penetrating TMC-1 more heavily on the NE side \citep[e.g.,][]{spezzano_2016,spezzano_2020}, a strong possibility given the steeper \nh\ gradient in this direction. It would be necessary to obtain precise measurements of the ISRF in the environment of TMC-1 to distinguish between these two scenarios. 

Regardless of the cause of asymmetry, we interpret the $\sim$4 $\times$ difference between the transverse velocity gradients of \Gi\ and \Gii\ as the deceleration of the material during gravitational inflow. The low-density material flows onto the filament and piles onto the higher density region towards the centre, causing the gas to slow down. The outer, lower density region is being traced by \Gi\ while the central, higher density region is being traced by \Gii. In the following discussion, we label the inner region the `spindle' of the filament, and the outer region the `flattened envelope'. We adopt a two-layer model where each layer is described by its density and velocity characteristics.

This does not necessarily imply that there is a sharp interface between the layers, but rather there could be a transition region, where the accretion-related kinematics in the flattened envelope change to fragmentation-related kinematics in the spindle. A sharp interface is more characteristic of shock compression than smooth gravitational inflow.
%
%
Higher spatial and spectral resolution observations would be needed to identify this transition region and characterize how rapidly the kinematics change. Similar interpretations have been presented for the filamentary star-forming regions Serpens South \citep{kirk_2013} and NGC 1333 \citep{MChen_2020}, where the authors suggest that accreted material is slowed and directed to flow along filaments. Our data are consistent with these interpretations. 

Previous studies of TMC-1 have already identified multiple velocity components using different chemical tracers. \citet{feher_2016} used a $k$-means clustering method with \vlsr\ maps derived from \amm\ (1, 1) and (2, 2) transitions. While the authors describe multiple velocity components in some lines of sight, their clustering method identifies regions based on position (RA), position (Dec)., and velocity of the primary component observed in \amm (1,1), and \hh\ column density. This investigation found four distinct kinematic regions of TMC-1, and their region named TMC-1F4 overlaps almost completely with our region of interest. They note that their TMC-1F4 has several regions with multiple velocity components, particularly on its northern edge where it overlaps with another one of their regions, TMC-1F2. Spatially, this bears resemblance to our analysis where \Gi\ and \Gii\ are found in the same region as TMC-1F4 while \Giii\ may correspond to the tail end of TMC-1F2 which extends into the upper half of the filament. As they did not include multiple velocity components in their clustering algorithm, these results are complementary to each other. 

\citet{Dobashi_2019} identify 21 velocity-coherent substructures toward TMC-1 by eye in position-velocity space of CCS emission, seven of which fall into our region of study at the south end of TMC-1. The authors interpret these structures as `fibres' \citep[][]{Hacar_2013}. The \vlsr\ of their two most prominent `fibres' correspond well with the \vlsr\ that we have identified for \Gi\ and \Gii. Their fibre 1 runs along the spine of the filament and has a \vlsr = 5.72\,\kms\ compared to 5.68\,\kms, the mean \vlsr\ of \Gii. Additionally, their fibre 2 has a \vlsr\ = 5.89\,\kms\ compared to the mean \vlsr\ = 5.94\,\kms\ of \Gi. The emission peak of \Gi\ is located near fibre 2. They identify a short `fibre' along the northern edge with a mean \vlsr = 6.16\,\kms\ which is very comparable in both velocity and location to the strongest emission in \Giii. The classification of \Giii\ in our physical picture is not clear given the signal-to-noise and spectral resolution of our data. It may be the southern extension of a separate physical component that is more prevalent in the northern half of TMC-1, or it may be a more red-shifted extension of the flattened envelope traced by \Gi. This may be resolved in a future study of TMC-1. The CSS molecule observed by \citet[][]{Dobashi_2019} and the \hcfn\ in this study are both carbon chains with rotational transitions that are excited at similar gas densities and should therefore trace similar regions of the filament. We find no evidence, however, of four additional velocity components in the south end of TMC-1.

Our approach and interpretation offer a slightly different picture for TMC-1 than the one suggested for L1495/B213 \citep[][]{Hacar_2013} and TMC-1 \citep{Dobashi_2019}, where velocity components are physically distinct intertwining `fibres' of gas. We also differ from \citet[][]{choi_2017}, who interpret a sharp change in the velocity field of N$_2$H$^+$ around the \hcfn\ emission peak as the collision of two distinct flows, causing the formation of a prestellar core at this location. 
In this model, TMC-1 does not have intertwining fibres, nor is it formed from two distinct, colliding flows. Rather, the \hcfn\ observed toward the TMC-1 filament traces two layers, each with their own distinct kinematics. The two layers blend and the gas transitions smoothly from \Gi\ to \Gii\ as it settles onto the spindle. Higher spatial and spectral resolution could also be used here to differentiate between our smooth transition interpretation and the discrete `fibres' picture.

\begin{figure*}
    \centering
    \includegraphics[width=0.9\linewidth]{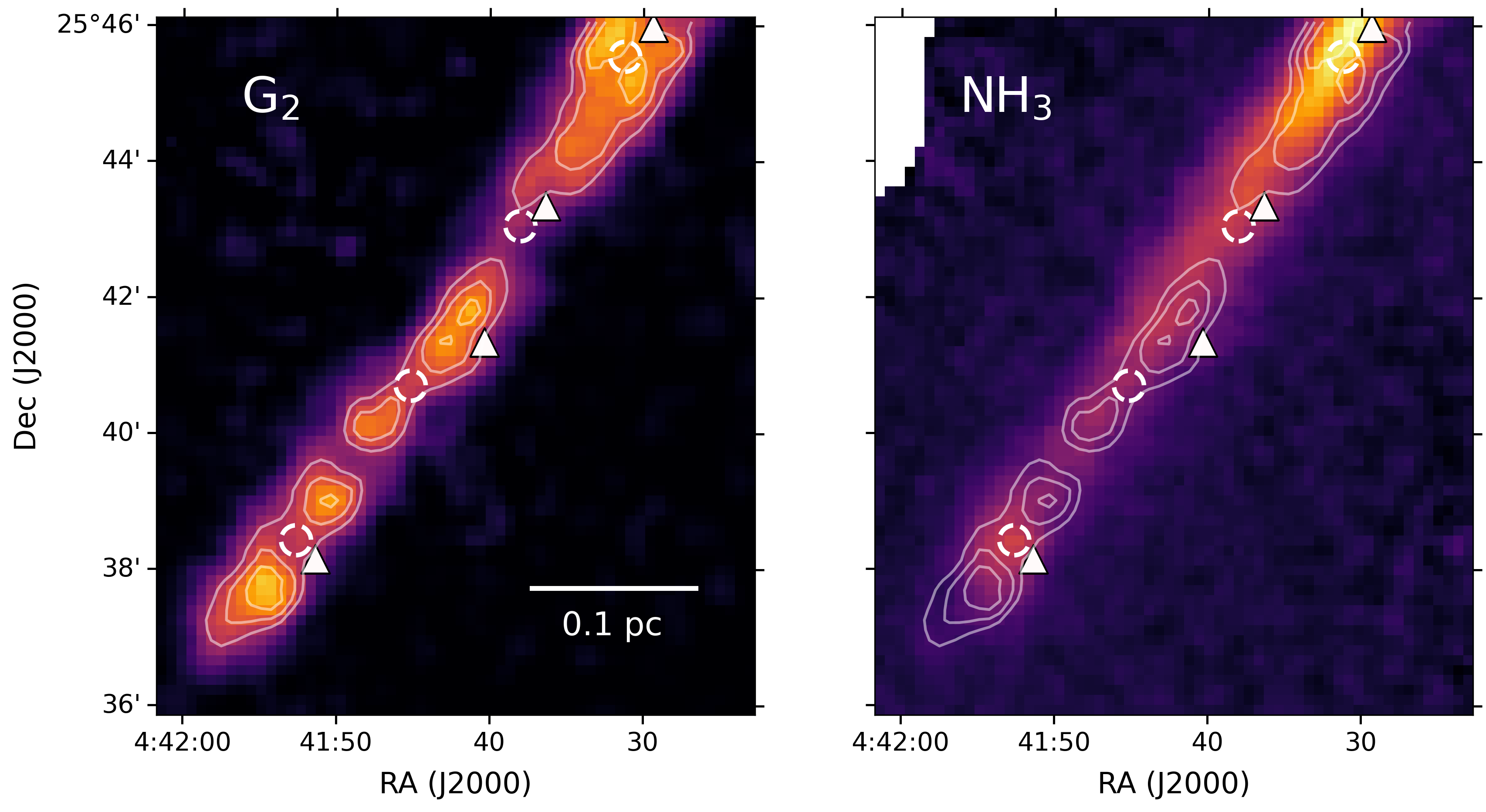}
    \caption{\textit{Left}: \Gii\ integrated intensity map. Dashed white circles show the locations of mid-points between \Gii\ emission peaks and solid white triangles show the locations of 850 $\mu$m JCMT peaks. Additionally, the faint white contours are 4.5, 6, and 7.5-$\sigma_{\text{bg}}$ with respect to the \Gii\ emission map. A 0.1\,pc scale bar is shown in the lower right corner. \textit{Right}: \amm\ (1,1) integrated intensity map. Same markers and contours as \textit{Left} panel. Note in particular the two \Gii\ peaks at the southern tip that are located on either side of the peak in the \amm.}
    \label{fig:clumps}
\end{figure*}

\subsection{Substructure in \Gii: Fragmentation}
\label{sec:clumps}

The \Gii\ integrated intensity, shown in Fig. \ref{fig:param_maps}, map shows structure that is not visible in either the \nh\ or the total \hcfn\ map (Fig. \ref{fig:all_three}). This may indicate that fragmentation is beginning to take shape in the spindle of TMC-1. We identify eight emission peaks along the length of the \Gii\ integrated intensity map using standard peak finding software in {\tt python}. They are regularly spaced with an average projected spacing for these peaks is 0.06 $\pm$ 0.01\,pc. 
TMC-1 may be inclined with respect to the plane of the sky.
If the filament were inclined within a range of 30$-$60\textdegree, the average deprojected separations would be 0.07$-$0.11\,pc. 

However, carbon chains like \hcfn\ deplete quickly from the gas phase at high densities, on timescales of $\sim$10$^5$\,yr at $n \sim 10^5$~\cc\ \citep[][]{suzuki_1992,sakai_2008,mcelroy_2013}. More chemically evolved cores tend to show decreased abundances of carbon-bearing species towards core centres \citep[e.g.,][]{suzuki_1992,tafalla_2002,tafalla_2004}.
If TMC-1 is beginning to fragment into cores, then we would expect \hcfn\ in the spindle to show stronger emission on either side of the forming cores due to depletion towards the higher density core centres. Additionally, \hcfn\ depletion at high densities could explain why the \Gii\ column density is less than half that of \Gi. In our envelope-spindle framework, \Gii\ emission is originating from a region of higher density deep within the TMC-1 filament. We find the midpoint between each pair of peaks in \Gii\ to estimate the location of putative cores and recalculate the spacing with the same inclination considerations. 
The projected spacing of the four mid-points is 0.12 $\pm$ 0.002\,pc while the deprojected spacings range from 0.14\,pc to 0.23\,pc if the inclination is somewhere between 30\textdegree\ and 60\textdegree. 
All lengths are calculated using an assumed distance of 140 pc (as noted in Section \ref{sec:intro}). 

Classical cylinder fragmentation theory predicts that interstellar turbulence induces regularly spaced density fluctuations which lead to the formation of prestellar cores in both gravitationally stable and unstable (trans-critical and super-critical) filaments. The mean core spacing for an infinitely long, isothermal cylinder in thermal equilibrium has been predicted to be 4\,$\times$ the filament diameter \citep{inutsuka_1992} or 5\,$\times$ the FWHM \citep{fischera_2012}. The FWHM of TMC-1 is $0.12 \pm 0.05$\,pc (found in Section \ref{sec:align}), giving a predicted spacing of $0.6 \pm 0.25$\,pc.
This vastly overpredicts the observed spacings even if extreme projection effects are considered.
However, the predicted spacing of cylindrical fragmentation theory is known to be a poor fit to fragmentation scales in observed filaments \citep[e.g.,][]{Zhang_2020} and in simulations \citep[e.g.,][]{clarke_2016}. Predicted fragmentation length scales vary when induced by different sources such as magnetic fields \citep{andre_2019}, turbulent accretion \citep{Clarke2017}, and geometrical perturbations \citep{Gritschneder2017}.

On the other hand, \citet[][]{coughlin_2020} have performed a theoretical study of an infinite, adiabatic, polytropic cylinder and its response to small perturbations. The fastest growing modes are separated by $\approx$\,1.5\,$\times$ the filament diameter which predicts a fragmentation length-scale of $0.18 \pm 0.075$\,pc in TMC-1. The average spacing we observe between \Gii\ mid-points is consistent with the predictions of \citet[][]{coughlin_2020} even with moderate inclination angle considerations. 

We also calculated the turbulent Jeans length following \citet{chandra_1951} and found it to be 0.11\,pc using a mean density of $4\times10^4$\,cm$^{-3}$. We note that, per our assumptions in the spindle-envelope model, the density of \Gii\ is likely higher than the average density of the filament. Thus, this calculation provides the upper bound for the Jeans length. Interestingly, this is similar to the predictions of \citet[][]{coughlin_2020}, but slightly underpredicts the spacings that are observed.
The spherical Jeans length has been compared with observations as a fragmentation length-scale in a variety of cloud complexes \citep{jackson_2010, kainulainen_2013, Zhang_2020}. 

The \nh\ data derived from {\it Herschel} does not show distinct, evenly spaced peaks indicating the presence of prestellar cores. \citet{Nutter_2008} analysed TMC-1 with 850\,$\mu$m continuum emission obtained from the SCUBA instrument \citep[][]{holland_1999} on the James Clerk Maxwell Telescope (JCMT). The 850 $\mu$m has higher resolution (14\arcsec), revealing several continuum emission peaks within the TMC-1 segment which we identify on Fig. \ref{fig:clumps}. Remarkably, three of the four 850\,$\mu$m peaks are located very near to three of the \Gii\ mid-points. One SCUBA peak did still fall between two neighbouring \Gii\ emission peaks but not near to any of the predicted mid-points. Instead, this SCUBA peak is found directly beside the historical cyanopolyyne peak. Similar to our envelope-spindle model, \citet{Nutter_2008} furthermore compare the 850\,$\mu$m profile and the profile of 160\,$\mu$m emission from the {\it Spitzer Space Telescope} \citep[][]{werner_2004} toward TMC-1. The authors identify a colder (8\,K), narrow component and a warmer (12\,K), broader component which they refer to as the `core' and `shoulder' of the filament, respectively.

In comparing to the \amm\ emission, the most compelling argument for our paired-peak method is seen by looking at the southern tip of TMC-1 in the right panel of Fig. \ref{fig:clumps}. The circles show \Gii\ \hcfn\ emission peaks which are found on either side of an increase in \amm, the peak of which matches closely to the paired-peak mid-point. We also plotted the mid-points with respect to the longitudinal velocity gradients in Fig. \ref{fig:slopes} and found that all of the mid-points align with either a local minima or local maxima.
Ordered, oscillatory patterns in longitudinal velocity profiles are predicted to have their peaks offset from density oscillations by $\lambda$/4 where $\lambda$ is the wavelength of the perturbations in the medium \citep{Hacar_2011, tafalla_2015}. This model is not in agreement with the TMC-1 data. Higher resolution may be needed to precisely identify how the midpoint locations related to the oscillatory patterns in the velocity profile.

\section{Conclusions}
\label{sec:conclusion}

We have analysed a narrow segment of the molecular gas filament TMC-1 in the TMC complex with the goal of studying the small-scale kinematics. We fit a three Gaussian velocity component model to \hcfn\ 9-8 line emission using the {\tt ROHSA} line fitting method. 
We compared our results to \amm (1,1) emission from the GAS collaboration and \hh\ column densities derived from dust continuum observed by {\it Herschel}. Our main findings and conclusions are summarised as follows:
\begin{itemize}
    \item[i)] The components \Gi\ and \Gii\ are both offset to the north-east of the spine of \nh\ by 14\,\arcsec\ and 10\,\arcsec\ respectively. \Gi\ has a transverse velocity gradient of $2.7 \pm 0.1$\,\kmspc, while the transverse gradient of \Gii\ is approximately one quarter the magnitude at $0.7 \pm 0.1$\,\kmspc.
    
    \item[ii)] We find that TMC-1 is thermally trans-critical and magnetic fields provide negligible support against gravity. We compare the effect of self-gravity against bulk-rotation as well as local turbulent motions and find that inflow due to self-gravity is likely the driver of the transverse gradient in \Gi.
    
    \item[iii)] We consider a two-layer model where \Gi\ is tracing infalling gas in a `flattened envelope' on the outside of the filament while \Gii\ is tracing denser gas in a central `spindle' of the cylindrical filament. This model does not imply a sharp interface between the layers but could instead have a transition region where accreting material from the flattened envelope slows down and settles onto the spindle.
    
    \item[iv)] We identify mid-points between pairs of emission peaks in \Gii. The spacing of these mid points are consistent with the theoretical fragmentation length-scale predicted by the polytropic, adiabatic cylinder  \citep{coughlin_2020}, and correlate well with peaks in 850 $\mu$m continuum peaks from SCUBA. Even though we do not see obvious prestellar cores in the \nh, \Gii\ could be tracing the early hints of fragmenting material deep within TMC-1.
    
\end{itemize}

In future studies, higher spatial resolution could give a more robust measurement of the offsets of \Gi\ and \Gii\ from the spine of the \nh\ as those differences are on the scale of the spatial resolution in this work. That offset will give a better measurement of the transition between \Gi\ and \Gii. Better velocity resolution will be important for investigating the relationship between \Gii\ mid-points and the longitudinal velocity gradient oscillations as there are theoretical predictions for the observed gradients around actively forming prestellar cores.

\section*{Acknowledgements}
The authors would like to thank the Anonymous Referee whose thorough feedback of this manuscript improved both the clarity of the writing and the robustness of the conclusions.

The authors would like to thank C-Y. Chen for their helpful comments while writing this paper. 

This work was begun through the Summer Undergraduate Research Program at the University of Toronto, partially funded by the Dunlap Institute for Astronomy \& Astrophysics. The Dunlap Institute is funded through an endowment established by the David Dunlap family and the University of Toronto. The University of Toronto operates on the traditional land of the Huron-Wendat, the Seneca,
and most recently, the Mississaugas of the Credit River;
we are grateful to have the opportunity to work on this
land. 
The Green Bank Observatory is a facility of the National Science Foundation operated under cooperative agreement by Associated Universities, Inc.

AG acknowledges support from the NSF via grants AST 2008101 and CAREER 2142300.
JDF and HK are supported by the National Research Council of Canada and by Natural Sciences and Engineering Research Council of Canada (NSERC) Discovery Grants.
AP acknowledges the support by the Russian Ministry of Science and
Education via the State Assignment Contract FEUZ-2020-0038.

This research made use of the following code packages: Astropy, a community-developed core Python package for Astronomy \citep[][]{astropy:2013, astropy:2018}; matplotlib \citep[][]{Hunter:2007}; Numpy \citep[][]{harris2020array}; and Scipy \citep[][]{2020SciPy-NMeth}.


\section*{Data Availability}
 
Data cubes and {\tt ROHSA} output are publicly available at the CADC via following link: \url{http://doi.org/10.11570/22.0080} \citep{tmc1}.



\bibliographystyle{mnras}
\bibliography{ref} 




\onecolumn 
\appendix

\section{Additional Tables and Figures}
\label{sec:appa}

\begin{figure*}
    \centering
    \captionsetup{justification=centering}
    \begin{subfigure}[b]{\linewidth}
        \includegraphics[width=0.95\linewidth]{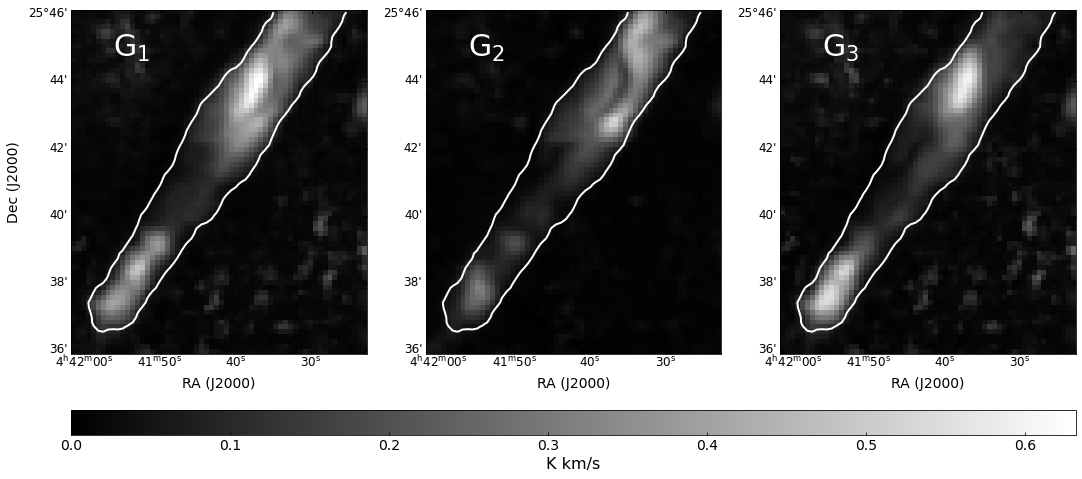}
    \end{subfigure}
    \begin{subfigure}[b]{\linewidth}
        \includegraphics[width=0.95\linewidth]{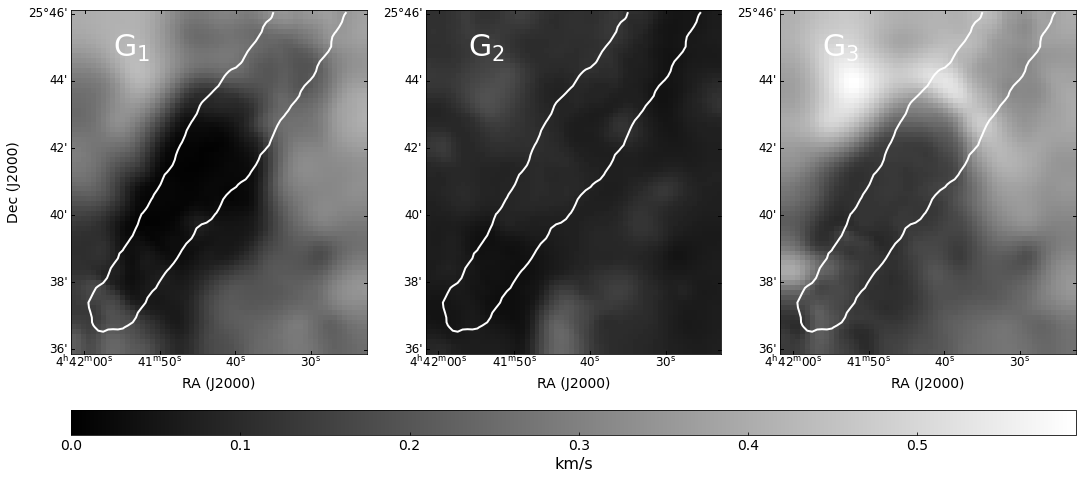}
    \end{subfigure}
    \begin{subfigure}[b]{\linewidth}
        \includegraphics[width=0.95\linewidth]{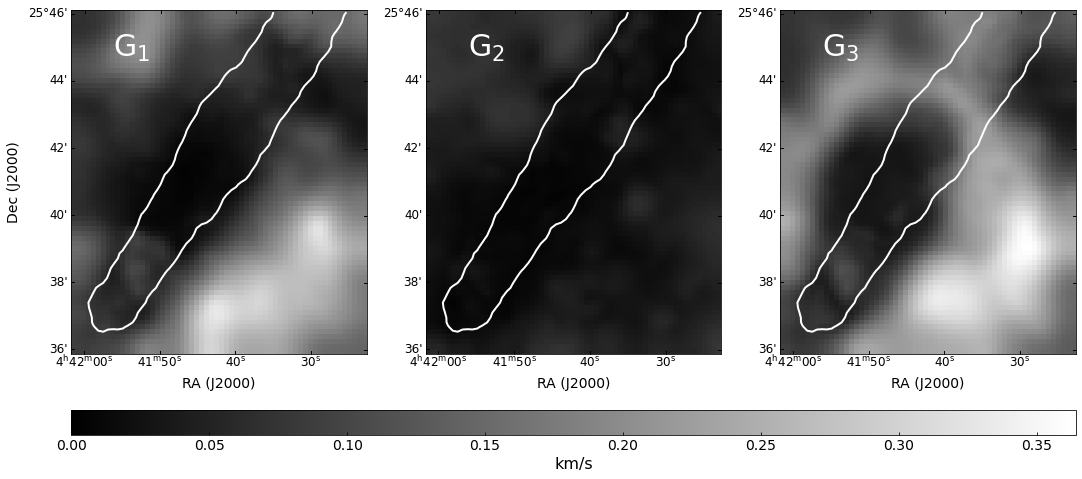}
    \end{subfigure}
    \caption{{\it Top}: Total uncertainties in the \Gi, \Gii, and \Giii\ integrated intensity maps. All three maps show increased uncertainty at the extrema of the filament.
    {\it Middle}: Total uncertainties in the \Gi, \Gii, and \Giii\ l.o.s. velocity maps. \Gi\ and \Gii\ both have low uncertainty, approximately equal to the velocity resolution. \Giii\ has much higher uncertainty on average.
    {\it Bottom}: Total uncertainties in the \Gi, \Gii, and \Giii\ velocity dispersion maps. The white contour shows the 5-$\sigma_{\text{bg}}$ boundary for the zeroth moment map.}
    \label{fig:uncert}
\end{figure*}

\begin{table}
    \centering
    \begin{tabular}{|c|c|}
        \hline
        {\bf N}  & {\bf $\chi^2$} \\
        \hline \hline
        1 & 3.08 \\
        2 & 2.16 \\
        3 & 1.55 \\
        4 & 1.61 \\
        5 & 2.07 \\
        \hline
    \end{tabular}
    \caption{Reduced $\chi^2$ value for $N$ gaussian components. $N=3$ was the best fit to the data.}
    \label{tab:Ng_chi2}
\end{table}

\begin{table}
    \centering
    \begin{tabular}{c|c|c}
        \hline
        {\bf Parameter Map} & {\bf Average Value} & {\bf Average Uncertainty}  \\
        \hline \hline
        $I_1$ & 0.72 K \kms\ &$\pm$ 0.20 K \kms\\
        $I_2$ & 0.38 K \kms\ &$\pm$ 0.15 K \kms\\
        $I_3$ & 0.22 K \kms\ &$\pm$ 0.17 K \kms\\
        $v_1$ & 5.94 \kms\ &$\pm$ 0.12 \kms\\
        $v_2$ & 5.68 \kms\ &$\pm$ 0.07 \kms\\
        $v_3$ & 6.20 \kms\ &$\pm$ 0.25 \kms\\
        $\sigma_1$ & 0.12 \kms\ &$\pm$ 0.05 \kms\\
        $\sigma_2$ & 0.09 \kms\ &$\pm$ 0.02 \kms\\
        $\sigma_3$ & 0.06 \kms\ &$\pm$ 0.10 \kms\\
        \hline
    \end{tabular}
    \caption{This table presents the average uncertainties calculated for the parameter maps of each velocity component that were calculated from noise injections and variations of the hyper-parameters in the {\tt ROHSA} fitting code. $v$ is \vlsr\ for each component.}
    \label{tab:uncerts}
\end{table}


\bsp	
\label{lastpage}
\end{document}